\documentclass[twocolumn,english]{revtex4}
\usepackage[T1]{fontenc}
\usepackage[latin9]{inputenc}
\usepackage{verbatim}
\usepackage{amstext}
\usepackage{amssymb}
\usepackage{graphicx}
\usepackage{esint}

\makeatletter
\@ifundefined{textcolor}{}
{%
 \definecolor{BLACK}{gray}{0}
 \definecolor{WHITE}{gray}{1}
 \definecolor{RED}{rgb}{1,0,0}
 \definecolor{GREEN}{rgb}{0,1,0}
 \definecolor{BLUE}{rgb}{0,0,1}
 \definecolor{CYAN}{cmyk}{1,0,0,0}
 \definecolor{MAGENTA}{cmyk}{0,1,0,0}
 \definecolor{YELLOW}{cmyk}{0,0,1,0}
}

\makeatother

\usepackage{babel}
\begin{document}
\global\long\def\ket#1{\left|#1\right\rangle }

\global\long\def\bra#1{\left\langle #1\right|}

\global\long\def\braket#1#2{\left\langle #1\left|#2\right.\right\rangle }

\global\long\def\ketbra#1#2{\left|#1\right\rangle \left\langle #2\right|}

\global\long\def\braOket#1#2#3{\left\langle #1\left|#2\right|#3\right\rangle }

\global\long\def\mc#1{\mathcal{#1}}

\global\long\def\nrm#1{\left\Vert #1\right\Vert }

\title{Quantum Equivalence and Quantum Signatures in Heat Engines}

\author{Raam Uzdin}

\author{Amikam Levy}

\author{Ronnie Kosloff}

\email{raam@mail.huji.ac.il}

\selectlanguage{english}%

\affiliation{Fritz Haber Research Center for Molecular Dynamics, Hebrew University
of Jerusalem, Jerusalem 91904, Israel}

\maketitle
Quantum heat engines (QHE) are thermal machines where the working
substance is quantum. In the extreme case the working medium can be
a single particle or a few level quantum system. The study of QHE
has shown a remarkable similarity with the standard thermodynamical
models, thus raising the issue what is quantum in quantum thermodynamics.
Our main result is thermodynamical equivalence of all engine type
in the quantum regime of small action. They have the same power, the
same heat, the same efficiency, and they even have the same relaxation
rates and relaxation modes. Furthermore, it is shown that QHE have
quantum-thermodynamic signature, i.e thermodynamic measurements can
confirm the presence of quantum coherence in the device. The coherent
work extraction mechanism enables power outputs that greatly exceed
the power of stochastic (dephased) engines.

\section{Introduction}

Thermodynamics emerged as a practical theory for evaluating the performance
of steam engines. Since then the theory proliferated and utilized
in countless systems and applications. Eventually, thermodynamics
became one of the pillars of theoretical physics. Amazingly, it survived
great scientific revolutions such as quantum mechanics and general
relativity. In fact, thermodynamics played a crucial role in the development
of these theories. Black body radiation led Planck and Einstein to
introduce energy quantization, and the second law of thermodynamics
led Bekenstein to discover the relation between the event horizon
area and black hole entropy and temperature. 

The Carnot efficiency is a manifestation of the second law in heat
engines, and it is universally valid. In addition all reversible engines
must operate at Carnot efficiency. Though very profound these principles
have limited practical value. First, real engines produce finite non
zero power and therefore cannot be reversible. Second, the performance
of real engine is more severely limited by heat leaks, friction and
heat transport. This led to the study of efficiency at maximal power
\cite{curzon75,novikov1958efficiency} and finite time thermodynamics
\cite{salamon01,AndresenFiniteTimeThermo2011}. In finite time thermodynamics
time was introduced classically using empirical heat transport models.
For microscopic systems the natural avenue to introduce time dynamics
is via quantum mechanics of open systems. The full quantum dynamics
may lead to new mechanism of extracting work or cooling. Alternatively,
it may lead to new microscopic heat leaks and friction-like mechanisms.

Quantum thermodynamics is the study of thermodynamic quantities such
as temperature, heat, work, and entropy in microscopic quantum systems
or even for a single particle. This includes dynamical analysis of
engines and refrigerators in the quantum regime \cite{alicki79,k24,k85,k152,k221,levy14,rahav12,allmahler10,linden10,correa14,mahler07b,skrzypczyk2014work,gelbwaser13,kolar13,alicki2014quantum,Nori2007QHE,lutz14,dorner2013extracting,dorner2012emergent,binder2014operational,DelCampo2014moreBang,gelbwaser2015Review,malabarba2014clock},
theoretical frameworks that takes into account single shot events
\cite{horodecki2013fundamental,verdal13}, and the study of thermalization
mechanisms \cite{mahlerbook,Eiset2012thermalizationInNature,Eisert2012probingTherm}. 

It was natural to expect that in the quantum regime new thermodynamic
effects will surface. However, quantum thermodynamic systems (even
with a single particle) show a remarkable similarity to macroscopic
system described by classical thermodynamic. When the baths are thermal
the Carnot efficiency limit is equally applicable for a small quantum
system \cite{alicki79,spohn78}. Even classical fluctuation theorems
hold without any alteration \cite{campisi09,campisi11,quan2008quantumFluctTheorem}. 

Is there really nothing new and profound in the thermodynamics of
small quantum system? In this work, we present thermodynamic behavior
that is purely quantum in its essence and has no classical counterpart. 

Recently some progress on the role of quantum coherence in quantum
thermodynamics has been made \cite{RudolphPRX_15,LostaglioRudolphCohConstraint}.
In addition, quantum coherence has been shown to quantitatively affect
the performance of heat machines \cite{mukamel12,scully03,scully2011quantum}.
In this work we associate coherence with a specific \textit{thermodynamic
effect} and relate it to a thermodynamic \textit{work extraction mechanism}. 

Heat engines can be classified by their different scheduling of the
interactions with the baths and the work repository. These types include
the four-stroke, two-stroke and the continuous engines (these engine
types will be described in more detail later on). The choice of engine
type is usually guided by convenience of analysis or ease of implementation.
Nevertheless, from a theoretical point of view, the fundamental differences
or similarities between the various engine types are still uncharted.
This is particularly true in the microscopic quantum regime. For brevity
we discuss engines but all our results are equally applicable to other
heat machines such as refrigerators and heaters.

Our first result is that in the limit of small engine action (weak
thermalization, and a weak driving field), all three engine types
are thermodynamically equivalent. The equivalence holds also for transients
and for states that are very far from thermal equilibrium. On top
of providing a thermodynamic unification limit for the various engine
types, this finding also establishes a connection to quantum mechanics
as it crucially depends on phase coherence and quantum interference.
In particular, the validity regime of the equivalence is expressed
in terms of $\hbar$.

Our second result concerns quantum-thermodynamic signatures. Let us
define a \textit{quantum signature} as a signal extracted from measurements
that unambiguously indicates the presence of quantum effects (e.g.
entanglement or interference). The Bell inequality for the EPR experiment
is a good example. A \textit{quantum-thermodynamic signature} is a
quantum signature obtained from measuring thermodynamic quantities.
We show that it is possible to set an upper bound on the work output
of a stochastic, coherence-free engine. Any engine that surpasses
this bound must have some level of coherence. Hence, work exceeding
the stochastic bound constitutes a quantum-thermodynamic signature.
Furthermore, we distinguish between a coherent work extraction mechanism
and a stochastic work extraction mechanism. This explains why in the
equivalence regime, coherent engines produce significantly more power
compared to the corresponding stochastic engine.

The equivalence derivation is based on three ingredients. First, we
introduce a multilevel embedding framework that enables the analysis
of all three types of engines in the same physical setup. Next, a
``norm action'' smallness parameter, $s$, is defined for engines
using Liouville space. The third ingredient is the symmetric rearrangement
theorem that is used to show why all three engine types have the same
thermodynamic properties despite the fact that they exhibit very different
density matrix dynamics. 

In section two we describe the main engine types, and introduce the
multilevel embedding framework. Next, in section three the multilevel
embedding and the symmetric rearrangement theorem are used to show
the various equivalence relation of different engine types. After
discussing the two fundamental work extraction mechanisms, in section
four we present and study the over-thermalization effect in coherent
heat engines. In section five we show a quantum thermodynamic signature
that separates quantum engines from stochastic engines. Finally, in
section six we conclude and discuss extensions and future prospects.

\section{Heat engines types and the multilevel embedding scheme}

Heat engines are either discrete such as the two stroke engines or
the four stroke Otto engine. Heat engines can also operate continuously
such as in turbines. Quantum analogues of all these engine types have
been studied\footnote{other types consist of small variations and combination of these types.}.
Here we present a theoretical framework where all three type of engines
can be embedded in a unified physical framework. This framework termed
``multilevel embedding'' is an essential ingredient in our theory
as it enables a meaningful comparison between different engine types.

\subsection{Heat and work}

A heat engine is a device that uses at least two thermal baths in
different temperatures to extract work. Loosely speaking, work is
the transfer of energy to a single degree of freedom. For example,
increasing the excitation number of an oscillator, increasing the
photon number in a specific optical mode (lasing) or increasing the
kinetic energy in a single predefined direction. ``Battery'' or
``flywheel'' are terms often used in this context of work storage
\cite{alicki13,hovhannisyan13}. We shall use the more general term
``work repository''. Heat, on the other hand, is energy spread over
multiple degrees of freedom. Close to equilibrium and in quasistatic
process, the heat is related to the temperature and to the effective
number of degrees of freedom (entropy S) via the well known relation
$dQ=TdS$.

In the elementary quantum heat engines the working substance comprises
of a single particle (or few at the most). Thus the working substance
cannot reach equilibrium on its own. Furthermore, excluding few non-generic
cases it is not possible to assign an equation of state that establishes
a relation between thermodynamic quantities when the substance is
in equilibrium. Nevertheless, QHE satisfy the second law and therefore
are also bounded by Carnot efficiency limit \cite{alicki79}.

Work strokes are characterized by zero contact with the baths and
an inherently time dependent Hamiltonian. The unitary evolution generated
by this Hamiltonian can change the energy of the system. On the other
hand, the von Neumann entropy and the purity, remain fixed (unitary
evolution at this stage). Hence the energy change of the system in
this case constitutes pure work. The system's energy change is actually
an energy exchange with the work repository.

When the system is coupled to a thermal bath and the Hamiltonian is
fixed in time, the bath can change the populations of the energy levels.
In steady state the system reaches a Gibbs state where the density
matrix has no coherences in the energy basis and the population of
the levels is given by: $p_{n,b}=e^{-\frac{E_{n}}{T_{b}}}/\sum_{n=1}^{N}e^{-\frac{E_{n}}{T_{b}}}$
where $N$ is the number of levels and $'b'$ stands for $'c'$ (cold)
or $'h'$ (hot). In physical models where the system thermalizes via
collision with bath particles a full thermalization can be achieved
in finite time \cite{gennaro2008entanglement,GennaroQuDit,rybar2012simulation,ziman2005description,RUswap}.
However, it is not necessary that the baths will bring the system
close to a Gibbs state for the proper operation of the engine. In
particular, maximal efficiency (e.g. in Otto engines) can be achieved
without full thermalization. Maximal power (work per cycle time) is
also associated with partial thermalization \cite{curzon75,esposito09}.
The definitive property of a thermal bath is its aspiration to bring
the system to a predefined temperature regardless of the initial state
of the system. The evolution in this stage does not conserve the eigenvalues
of the density matrix of the system, and therefore not only energy
but entropy as well is exchanged with the bath. Therefore, the energy
exchange in this stage is considered as heat.

In contrast to definitions of heat and work that are based on the
derivative of the internal energy \cite{alicki79,k281,anders2013thermodynamics},
our definitions are obtained by energy balance when coupling only
one element (bath or external field) at a time. As we shall see later
on, in some engine types several agents change the internal energy
simultaneously. Even in this case, this point of view of heat and
work will still be useful for obtaining consistent and physical definitions
of heat and work.

\subsection{The three engine types}

There are three core engine types that operate with two thermal baths:
four-stroke engine, two-stroke engine, and a continuous engine. A
stroke is a time segment where a certain operation takes place, for
example thermalization or work extraction. Each stroke is a $CP$
map and therefore the one-cycle evolution operator of the engine is
also a $CP$ map (since it is a product of $CP$ maps). For the extraction
of work it is imperative that some of the stroke propagators do not
commute \cite{k258}.

Otto engines and Carnot engines are examples of four stroke engines.
The simplest quantum four-stroke engine is the two-level Otto engine
shown in Fig. 1a. In the first stroke only the cold bath is connected.
Thus, the internal energy changes are associated with heat exchange
with the cold bath. The expansion and compression of the levels are
fully described by a time-dependent Hamiltonian of the form $H(t)=f(t)\sigma_{z}$
(the baths are disconnected at this stage). In the second stroke,
work is consumed in order to expand the levels, and in the fourth
stroke work is produced when levels revert to their original values.
There is a net work extraction since the populations in stages II
and IV are different. 
\begin{figure}
\begin{centering}
\includegraphics[width=8.6cm]{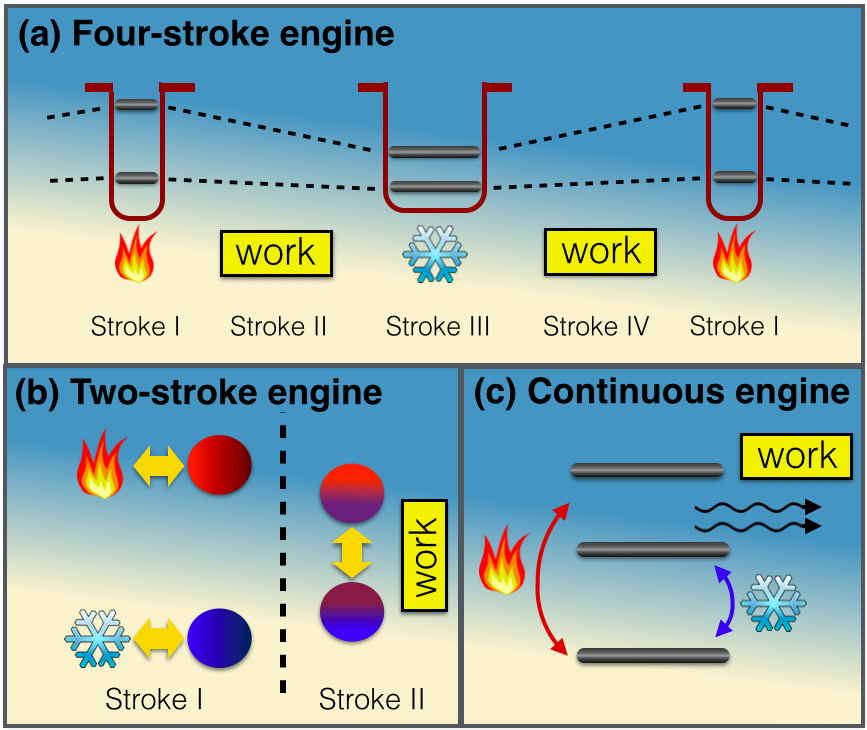}
\par\end{centering}

\centering{}\label{fig1}\protect\caption{(a) A two-level scheme of a four-stroke engine. (b) Two-particle scheme
of a two stroke engine. (c) A three-level scheme of a continuous engine.}
\end{figure}
As we shall see later on, other unitary operations are more relevant
for quantum equivalence of heat engines. Nevertheless, this particular
operation resembles the classical expansion and compression of classical
engines. The work is the energy exchanged with the system during the
unitary stages: $W=W_{II}+W_{IV}=(\left\langle E_{3}\right\rangle -\left\langle E_{2}\right\rangle )+(\left\langle E_{5}\right\rangle -\left\langle E_{4}\right\rangle )$.
We will consider only energy expectation values for two main reasons.
First, investigations of work fluctuations revealed that quantum heat
engines follow classical fluctuation laws \cite{campisi14} and we
search for quantum signatures in heat engines. The second reason is
that in our view the engine should not be measured during operation.
The measurement protocol used in quantum fluctuation theorems \cite{campisi09,campisi11,campisi14},
eliminates the density matrix coherences. These coherences have a
critical component in the equivalence and quantum signature we study
in this paper. Thus, although we frequently calculate work per cycle,
the measured quantity is the cumulative work and it is measured only
at the end of the process. The averaged quantities are obtained by
repeating the full experiment many times. Engines are designed to
perform a task and we assume that this completed task is the subject
of measurement. The engine internal state are not measured. 

The heat per cycle taken from the cold bath is $Q_{c}=\left\langle E_{2}\right\rangle -\left\langle E_{1}\right\rangle $
and the heat taken from the hot bath is $Q_{h}=\left\langle E_{4}\right\rangle -\left\langle E_{3}\right\rangle $.
In steady state the average energy of the \textit{system} returns
to its initial value after one cycle\footnote{This is, of course, not true for the work repository.}
so that $\left\langle E_{5}\right\rangle =\left\langle E_{1}\right\rangle $.
From this it follows immediately that $Q_{c}+Q_{h}+W=0$, i.e. the
first law of thermodynamics is obeyed. There is no instantaneous energy
conservation of \textit{internal} energy, as energy may be temporarily
stored in the interaction field or in the work repository. 

In the two-stroke engine shown in Fig 1b the engine consists of two
parts (e.g. two qubits) \cite{AllahverdyanOptDualStroke}. One part
may couple only to the hot bath and the other may couple only to the
cold bath. In the first stroke both parts interact with their bath
(but do not necessarily reach equilibrium). In the second unitary
stroke the two engine parts are disconnected from the baths and are
coupled to each other. They undergo a mutual unitary evolution and
work is extracted in the process. 

In the continuous engine shown in Fig. 1c the two baths and the external
interaction field are connected continuously. For example, in the
three level laser system shown in Fig 1c the laser light represented
by $\mc H_{w}(t)$ generates stimulated emission that extracts work
from the system. This system was first studied in thermodynamics context
in \cite{scovil59}, while a more comprehensive analysis of the system
was given in \cite{k122}. It is imperative that the external field
is time-dependent. If it is time-independent the problem becomes a
pure heat transport problem where $Q_{h}=-Q_{c}\neq0$. In heat transport
the interaction field merely ``dresses'' the level so that the baths
see a slightly modified system. The Lindblad generators are modified
accordingly and heat flows without extracting or consuming work \cite{levy214}.
Variations on these engine types may emerge due to realization constraints.
For example, in the two stroke engine the baths may be continuously
connected. This variation and others can still be analyzed using the
tools presented in this paper.

\subsection{Efficiency vs. work and heat}

Since the early days of Carnot, efficiency received considerable attention
for two main reasons. First, this quantity is of great interest from
both theoretical and practical points of view. The second reason is
that efficiency satisfies a universal bound that is independent of
the engine details. The Carnot efficiency bound is a manifestation
of the second law of thermodynamics. Indeed, for Markovian bath dynamics
it was shown that quantum heat engines cannot exceed the Carnot efficiency
\cite{alicki79}. Recently, a more general approach based on a fluctuation
theorem for QHE showed that the Carnot bound still holds for quantum
engines \cite{campisi14}. Studies in which higher than Carnot efficiency
are reported \cite{scully03}, are interesting but they use non-thermal
baths and therefore not surprisingly deviate from results derived
in the thermodynamic framework that deals with thermal baths. For
example, an electric engine is not limited to Carnot efficiency since
its power source is not thermal. Although, the present work has an
impact on efficiency as well, we focus on work and heat separately
in order to unravel quantum effects. As will be exemplified later,
in some elementary cases these quantum effects do not influence the
efficiency.

\subsection{Bath description and Liouville space}

The dynamics of the working fluid (system) interacting with the heat
baths is described by Lindblad-Gorini-Kossakowski-Sudarshan (LGKS)
master equation for the density matrix \cite{breuer,lindblad76,gorini276}:
\begin{equation}
d_{t}\rho=L(\rho)=-i[H_{s},\rho]+\sum_{k}A_{k}\rho A_{k}^{\dagger}-\frac{1}{2}A_{k}^{\dagger}A_{k}\rho-\frac{1}{2}\rho A_{k}^{\dagger}A_{k},\label{eq: Lind Hil}
\end{equation}
where the $A_{k}$ operators depend on the temperature, relaxation
time of the bath, system bath coupling, and also on the system Hamiltonian
$H_{s}$ \cite{breuer}. This form already encapsulates within the
Markovian assumption of no memory. The justification for these equations
arises from a ``microscopic derivation'' in the weak system bath
coupling limit \cite{davies74}. In this derivation a weak interaction
field couples the system of interest to a large system (the bath)
with temperature $T$. This interaction brings the system into a Gibbs
state at temperature $T$. The Lindblad thermalization operators $A_{k}$
used for the baths are described in the next section

Equation (\ref{eq: Lind Hil}) is a linear equation so it can always
be rearranged into a vector equation. Given an index mapping $\rho_{N\times N}\to\ket{\rho}_{1\times N^{2}}$
the Lindblad equation now reads:
\begin{equation}
id_{t}\ket{\rho}=(\mc H_{s}+\mc L)\ket{\rho},\label{eq: Lind Lio}
\end{equation}
where $\mc H_{s}$ is an Hermitian $N^{2}\times N^{2}$ matrix that
originates from $H_{s}$, and $\mc L$ is a non Hermitian $N^{2}\times N^{2}$
matrix that originates from the Lindblad evolution generators $A_{k}$.
This extended space is called Liouville space \cite{mukamel1995principles}.
In this paper we will use calligraphic letters to describe operators
in Liouville space and ordinary letters for operators in Hilbert space.
For states, however, $\ket A$ will denote a vector in Liouville space
formed from $A_{N\times N}$ by ``vec-ing'' $A$ into a column in
the same procedure $\rho$ is converted into $\ket{\rho}$. A short
review of Liouville space and some of its properties is given in appendix
II.

While the Lindblad description works very well for sufficiently long
times it fails for very short times where some of the approximation
breaks down. In scales where the bath still has a memory of the system's
past states, the semi group property of the Lindblad equation no longer
holds: $\ket{\rho(t+t')}\neq e^{-i(\mc H_{s}+\mc L)(t-t')}\ket{\rho(t')}$.
This will set a cutoff limit for the validity of the engine types
equivalence in the Markovian approximation.

Next we introduce the multilevel embedding scheme that enables us
to discuss various heat engines in the same physical setup.

\begin{figure}
\begin{centering}
\includegraphics[width=8.6cm]{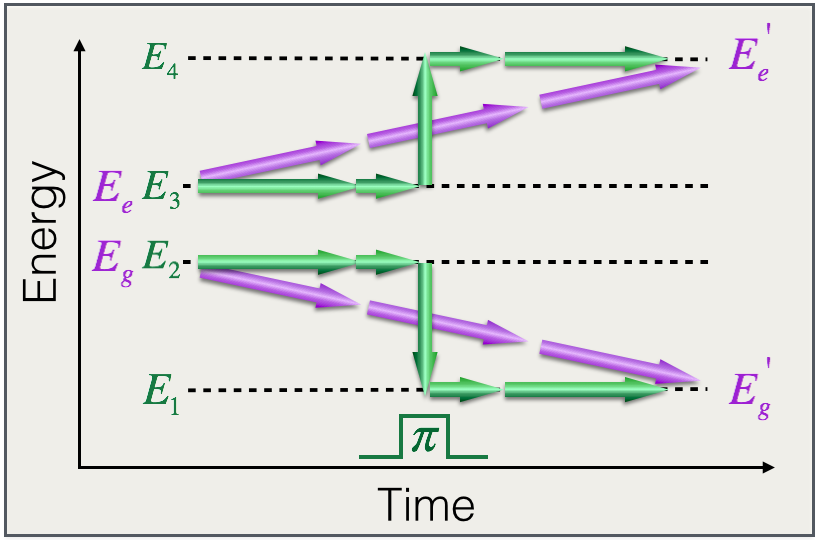}
\par\end{centering}

\label{fig2}\protect\caption{In the standard two-level Otto engine there are two-level $E_{g,e}$
(purple) that change in time to $E_{g,e}'$. In the multilevel embedding
framework, the levels ($E_{1-4}$) are fixed in time (black dashed
lines) but a time dependent field ($\pi$ pulse, swap operation) transfer
the population (green) to the other levels. For a swap operation the
two schemes lead to the same final state and therefore are associate
with the same work. Nonetheless, the multilevel scheme is more general
since for weaker unitary transformation (instead of the $\pi$ pulse),
coherences are generated. We show that this type of coherences can
significantly boost the power output of the engine.}
\end{figure}

\subsection{Multilevel embedding}

Let the working substance of the quantum engine be an $N$-level system.
These levels are fixed in time (i.e. they do not change as in Fig.
1a). For simplicity, the levels are non degenerate. We divide the
energy levels into a cold manifold and a hot manifold. During the
operation of the engine the levels in the cold manifold interact only
with the cold bath, and the hot manifold interact only with the hot
bath. Each thermal coupling can be turned on and off as a function
of time but the aliasing of a level to a manifold does not change
in time. 

If the manifolds do not overlap the hot and cold thermal operations
commute and they can be applied at the same time or one after the
other. The end result will be the same. Nevertheless, our scheme also
includes the possibility that one level appears in both manifold.
This is the case for the three-level continuous engine shown in Fig
1c. For simplicity, we exclude the possibility of more than one mutual
level. If there are two or more overlapping levels there is an inevitable
heat transport in steady state from the hot bath to the cold bath
even in the absence of an external field that extract work. In the
context of heat engines this can be interpreted as heat leak. This
``no field - no transport'' condition holds for many engines studied
in the literature. Nonetheless, this condition is not a necessary
condition for the validity of our results. 

This manifold division seems sensible for the continuous engine and
even for the two stroke engine in Fig 1b, but how can it be applied
to the four stroke engine shown in Fig. 1a? The two levels interact
with both baths and also change their energy value in time contrary
to the assumption of fixed energy levels. Nevertheless, this engine
is also incorporated in the multilevel embedding framework\textit{.}
Instead of two-levels as in Fig. 1a consider the four-level system
shown in the dashed green lines in Fig 2. 

\begin{figure}
\begin{centering}
\includegraphics[width=8.6cm]{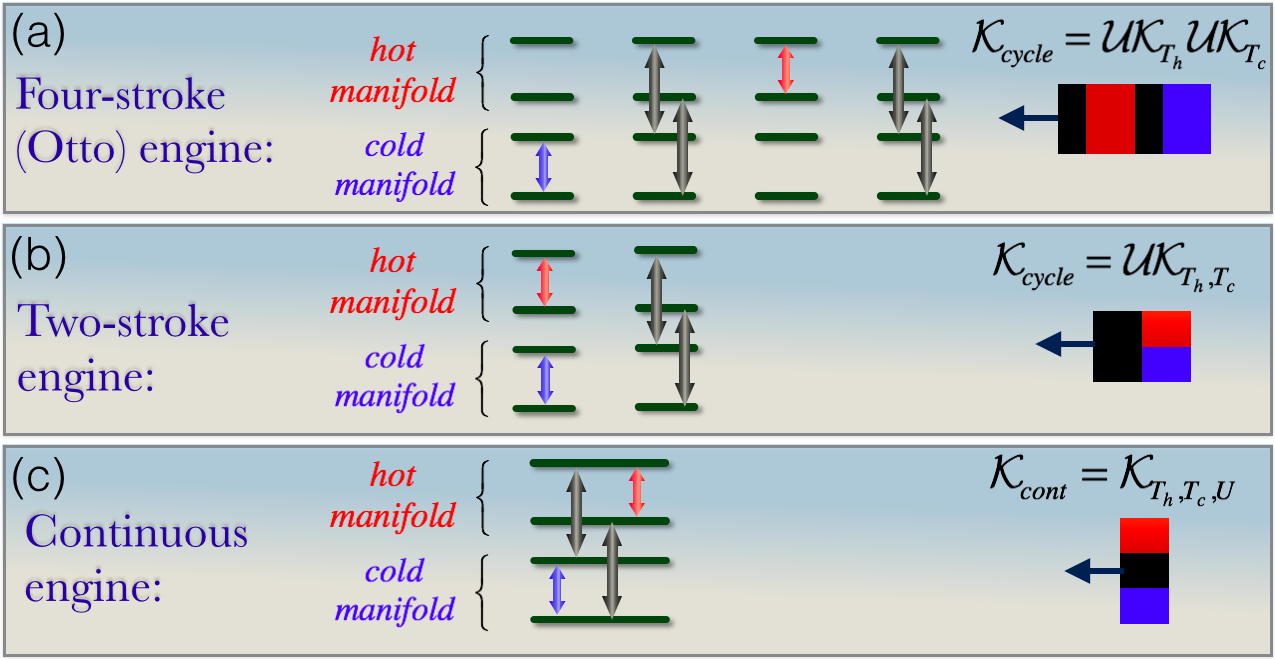}
\par\end{centering}

\label{fig3}\protect\caption{Representation of the three types of engines in the multilevel embedding
framework. In this scheme the different engine types differ only in
the order of coupling to the baths and work repository. Since the
interactions and energy levels are the same for all engine types,
a meaningful comparison of performance becomes possible. }
\end{figure}
Initially, only levels 2 and 3 are populated and coupled to the cold
bath (2 \& 3 are in the cold manifold). In the unitary stage an interaction
Hamiltonian $H_{swap}$ generates a full swap of populations and coherence
according to the rule $1\leftrightarrow2,3\leftrightarrow4$. Now
levels 1 and 4 are populated and 2 and 3 are empty. Therefore, this
system fully simulates the expanding level engine shown in Fig. 1a.
At the same time this system satisfies the separation into well defined
time-independent manifolds as defined in the multilevel embedding
scheme. 

The full swap used to embed the traditional four-stroke Otto engine,
is not mandatory and other unitary operations can be applied. This
extension of the four-stroke scheme is critical for our work since
the equivalence of engines appear when the unitary operation is fairly
close to the identity transformation.

Figure 3 shows how the three engine types are represented in the multilevel
embedding scheme. The advantage of the multilevel scheme now becomes
clear. All three engine types can be described in the same physical
system with the same baths and the same coupling to external fields
(work extraction). The engine types differ only in the order of the
coupling to the baths and to the work repository. While the thermal
operations commute if the manifolds do not overlap, the unitary operation
never commutes with the thermal strokes. 

On the right of Fig. 3, we plotted a ``brick'' diagram for the evolution
operator. Black stands for unitary transformation generated by some
external field, while blue and red stand for hot and cold thermal
coupling, respectively. When the bricks are on top of each other it
means that they operate simultaneously. Now we are in position to
derive the first main results of this paper: the thermodynamic equivalence
of engine types in the quantum regime.

\section{Continuous and stroke engine equivalence}

We first discuss the equivalence of continuous and four-stroke engines.
Nevertheless, all the argument are valid for the two-stroke engines
as well, as explained later on. Although our results are not limited
to a specific engine model it will be useful to consider the simple
engine shown in Fig. 4 . We will use this model to highlight a few
points and also for numerical simulations. The Hamiltonian part of
the system is:
\begin{equation}
H_{0}+\cos(\omega t)H_{w},\label{eq: H(t) num}
\end{equation}
where $H_{0}=-\frac{\Delta E_{h}}{2}\ketbra 11-\frac{\Delta E_{c}}{2}\ketbra 22+\frac{\Delta E_{c}}{2}\ketbra 33+\frac{\Delta E_{h}}{2}\ketbra 44$,
$H_{w}=\epsilon(t)\ketbra 12+\epsilon(t)\ketbra 34+h.c.$ and $\omega=\frac{\Delta E_{h}-\Delta E_{c}}{2}$. 

The driving frequency, that couples the system to the work repository
is in resonance with the top and bottom energy gaps. The specific
partitioning into hot and cold manifolds was chosen so that only one
frequency (e.g. a single laser) is needed for implementing the system
instead of two. 

\begin{figure}
\begin{centering}
\includegraphics[width=8.6cm]{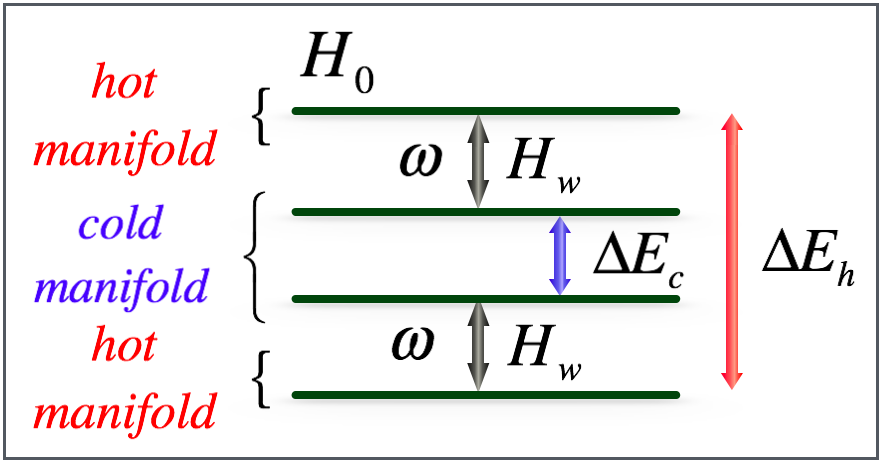}\protect\caption{Illustration of the engine used in the numerical simulation. By changing
the time order of the coupling to $H_{w}$ and to thermal baths, all
three type of engines can realized in the model. }

\par\end{centering}

\end{figure}

We assume that the Rabi frequency of the drive $\epsilon$ is smaller
than the decay time scale of the baths $\epsilon\ll\gamma_{c},\gamma_{h}$.
Under this assumption the dressing effect of the driving field on
the system-bath interaction can be ignored. It is justified, then,
to use ``local'' Lindblad operators obtained in the absence of a
driving field \cite{plenio10,levy214}. For plotting purposes (reasonable
duty cycle) in the numerical examples we will often use $\epsilon=\gamma_{c}=\gamma_{h}$.
While this poses no problem for stroke engines realizations, for experimental
demonstration of equivalence with continuous engines one has to increase
the duty cycle so that $\epsilon\ll\gamma_{c},\gamma_{h}$. That is,
the unitary stage should be made longer but with weaker driving field.

The Lindblad equation is given by (\ref{eq: Lind Hil}) with the Hamiltonian
(\ref{eq: H(t) num}) and with the following Lindblad operators:
\begin{eqnarray*}
A_{1} & = & \sqrt{\gamma_{h}}e^{-\frac{\Delta E_{h}}{2T_{h}}}\ketbra 41,\\
A_{2} & = & \sqrt{\gamma_{h}}\ketbra 14,\\
A_{3} & = & \sqrt{\gamma_{c}}e^{-\frac{\Delta E_{c}}{2T_{c}}}\ketbra 32,\\
A_{4} & = & \sqrt{\gamma_{c}}\ketbra 23.
\end{eqnarray*}
In all the numerical simulation we use $\Delta E_{h}=4$, $\Delta E_{c}=1$,
$T_{h}=5$, $T_{c}=1$. The interaction with the baths or with work
repository can be turned on and off at will. 

Starting with the continuous engine, we choose a unit cell that contains
exactly $6m$ ($m$ is an integer) complete cycles of the drive ($\tau_{d}=2\pi/\omega$)
so that $\tau_{cyc}=6m\tau_{d}$. The difference between the engine
cycle time and the cycles of the external drive, will become clear
in stroke engines (also the factor of six will be clarified). 

For the validity of the secular approximation used in the Lindblad
microscopic derivation \cite{breuer}, the evolution time scale must
satisfy $\tau\gg\frac{2\pi}{min(\Delta E_{h},\Delta E_{c})}$. Therefore
$m$ must satisfy $m\gg\frac{\omega}{min(\Delta E_{h},\Delta E_{c})}$.
Note that if the Lindblad description is obtained from a different
physical mechanism (e.g. thermalizing collisions) then this condition
is not required.

Next we transform to the interaction picture (denoted by tilde) using
the transformation $\mc U=e^{-i\mc H_{0}t}$, and perform the rotating
wave approximation (RWA) by dropping terms oscillating in frequency
of $2\omega$. For the RWA to be valid the amplitude of the field
must satisfy $\epsilon\ll\omega$. The resulting Liouville space super
Hamiltonian is:

\begin{equation}
\tilde{\mc H}=\mc L_{c}+\mc L_{h}+\frac{1}{2}\mc H_{w}.\label{eq: H ip-1}
\end{equation}
Note that $\mc L_{h,c}$ were not modified by the transformation to
the rotating system since $[\mc L_{h,c},\mc H_{0}]=0$ in the microscopic
derivation\footnote{This can be seen by following the derivation in \cite{breuer} and
using formalism introduced in \cite{machnes14}.}. Now that we have established a regime of validity and the super
Hamiltonian that governs the system, we can turn to the task of transforming
from one engine type to other types and study what properties change
in this transformation. The engine type transformation is based on
the Strang decomposition \cite{StrangDecompError2000jahnke} for two
non commuting operators $\mc A$ and $\mc B$ (the operators may not
be Hermitian): 
\begin{equation}
e^{(\mc A+\mc B)dt}=e^{\frac{1}{2}\mc Adt}e^{\mc Bdt}e^{\frac{1}{2}\mc Adt}+O(s^{3})\cong e^{\frac{1}{2}\mc Adt}e^{\mc Bdt}e^{\frac{1}{2}\mc Adt},\label{eq: Strang}
\end{equation}
where $s=(\nrm{\mc A}+\nrm{\mc B})dt$ must be small for the expansion
to be valid. $\nrm{\mc A}$ is the spectral norm (or operator norm)
of $\mc A$, and it is the largest singular value of $\mc A$, $\nrm{\mc A}=max\sqrt{eig(\mc A\mc A^{\dagger})}$\cite{roger1994topics}.
For Hermitian operators with eigenvalues $\lambda_{\mc A,i}$ the
spectral norm is $max(\left|\lambda_{\mc A,i}\right|)$. In Appendix
I we derive the condition $s\ll\frac{1}{2}\hbar$ for the validity
of (\ref{eq: Strang}). We will use the symbol $\cong$ to denote
equality with correction $O(s^{3})$. Let the evolution operator of
the continuous engine over the chosen cycle time $\tau_{cyc}=6m\tau_{d}$
be:
\begin{equation}
\tilde{\mc K}^{\text{cont}}=e^{-i\tilde{\mc H}\tau_{cyc}}.
\end{equation}
By first splitting $\mc L_{c}$ and then splitting $\mc L_{h}$ we
get:
\begin{eqnarray}
\tilde{\mc K}^{\text{4 stroke}} & = & e^{-i(3\mc L_{c})\frac{\tau_{cyc}}{6}}e^{-i(\frac{3}{2}\mc H_{w})\frac{\tau_{cyc}}{6}}e^{-i(3\mc L_{h})\frac{\tau_{cyc}}{3}}\nonumber \\
 & \times & e^{-i(\frac{3}{2}\mc H_{w})\frac{\tau_{cyc}}{6}}e^{-i(3\mc L_{c})\frac{\tau_{cyc}}{6}}.\label{eq: K 4s}
\end{eqnarray}
\begin{figure}
\includegraphics[width=8.6cm]{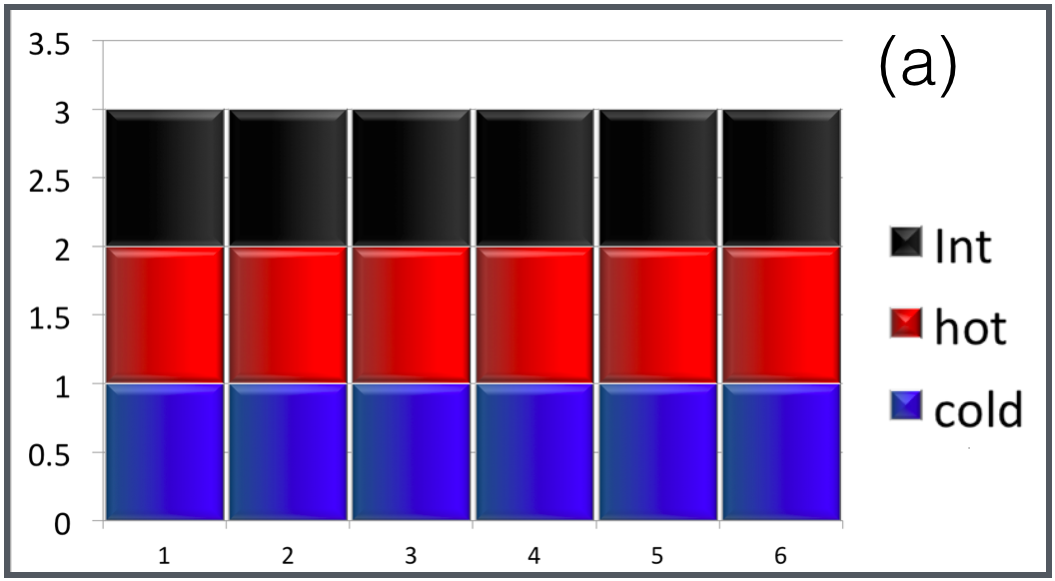}

\includegraphics[width=8.6cm]{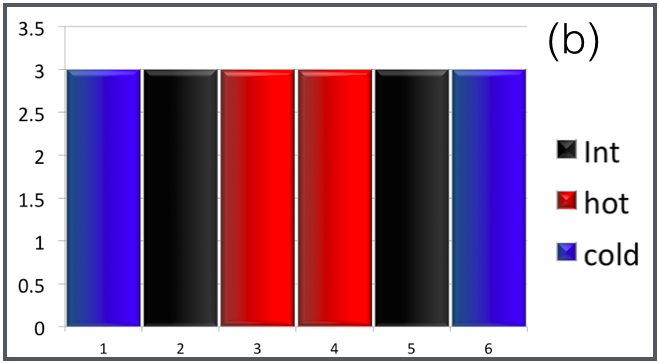}

\includegraphics[width=8.6cm]{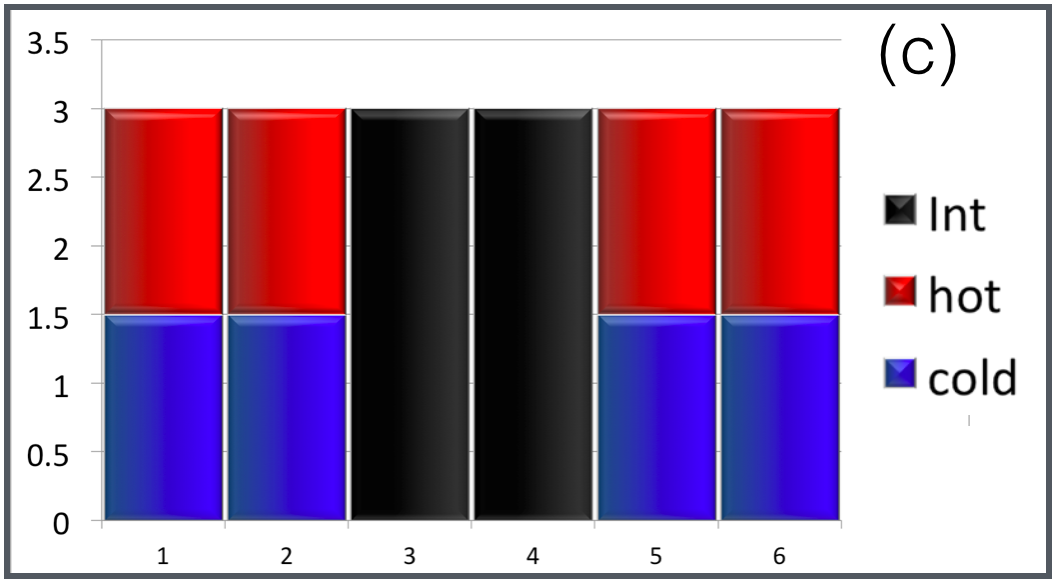}

\includegraphics[width=8.6cm]{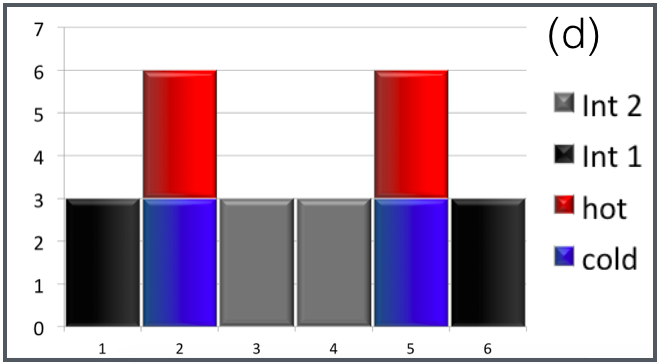}

\protect\caption{\label{fig5}(color online) Illustration of the splitting of the evolution
operator of the continuous engine (a), into a four-stroke engine (b),
a two-stroke engine (c), and a two-field four-stroke engine (d). The
horizontal axis corresponds to time as before. The size of the brick
corresponds to the strength of coupling to the work repository or
to the baths. The symmetric rearrangement theorem ensures that in
the limit of small action, any rearrangement that is symmetric with
respect to the center, and conserves the area of each color does not
change the total power and heat. }
\end{figure}
Note that the system is periodic so the first and last stages are
two parts of the same thermal stroke. Consequently (\ref{eq: K 4s})
describes an evolution operator of a four stroke engine, where the
unit cell is symmetric. This splitting is illustrated in Fig. 5a and
Fig. 5b. There are two thermal strokes and two work strokes that together
constitute an evolution operator that describes a four-stroke engine.
The cumulative evolution time as written above is $(m+m+2m+m+m)\tau_{d}=6m\tau_{d}=\tau_{cyc}$.
Yet, to maintain the same cycle time as chosen for the continuous
engine, the coupling to the baths and field were multiplied by three.
In this four-stroke engine each thermal or work stroke operates in
total only a third of the cycle time compared to the continuous engine.
Hence, the coupling must be three times larger in order to generate
the same evolution. 

By virtue of the Strang decomposition $\tilde{\mc K}^{\text{4 stroke}}\cong\tilde{\mc K}^{\text{cont}}$
if $s\ll1$. The \textit{action parameter} $s$ of the engine is defined
as $s=\intop_{-\tau_{cyc}/2}^{\tau_{cyc}/2}\nrm{\tilde{\mc H}}dt=(\frac{1}{2}\nrm{\mc H_{w}}+\nrm{\mc L_{h}}+\nrm{\mc L_{c}})\tau_{cyc}$.
Although we used $\hbar=1$ units, note that $s$ has dimensions of
$\hbar$ and therefore the relation $\tilde{\mc K}^{\text{4 stroke}}\cong\tilde{\mc K}^{\text{cont}}$
holds only when the engine action is small compared to $\hbar$. This
first appearance of a quantum scale will be discussed later on.

\subsection{Dynamical aspect of the equivalence}

Before discussing the thermodynamics properties of the engine we point
out that $\tilde{\mc K}^{\text{4 stroke}}\cong\tilde{\mc K}^{\text{cont}}$
has two immediate important consequences. First both engines have
the same steady state solution over one cycle $\ket{\tilde{\rho}_{s}}$:
\begin{eqnarray}
\tilde{\mc K}^{\text{4 stroke}}(\tau_{cyc})\ket{\tilde{\rho}_{s}}\cong\tilde{\mc K}^{\text{cont}}(\tau_{cyc})\ket{\tilde{\rho}_{s}} & = & \ket{\tilde{\rho}_{s}},\\
(\mc L_{c}+\mc L_{h}+\frac{1}{2}\mc H_{w})\ket{\tilde{\rho}_{s}} & = & 0.\label{eq: steady state}
\end{eqnarray}
At time instances that are not an integer multiples $\tau_{cyc}$,
the states of the engines will differ significantly ($O(s^{1})$)
since $\tilde{\mc K}^{\text{4 stroke}}(t<\tau_{cyc})\neq\tilde{\mc K}^{\text{cont}}(t<\tau_{cyc})$.
That is, the engines are still significantly different from each other.
The second consequence is that the two engines have the same transient
modes as well. When monitored at multiple of $\tau_{cyc}$ both engines
will have the same relaxation dynamics to steady state if they started
from the same initial condition. In the reminder of the paper, when
the evolution operator is written without a time tag, it means we
consider the evolution operator of a complete cycle.

\subsection{\label{sub: Thermo equiv}Thermodynamic aspect of the equivalence}

The equivalence of the one cycle evolution operators of the two engines
does not immediately imply that the engines are thermodynamically
equivalent. Generally, in stroke engines the heat and work depend
on the dynamics of the state inside the cycle which is very different
($O(s^{1})$) from the constant state of the continuous engine. However,
in this section we show that all thermodynamics properties are equivalent
in both engines up to $O(s^{3})$ corrections, similarly to the evolution
operator. We start by evaluating the work and heat in the continuous
engine. By considering infinitesimal time elements where $\mc L_{c},\mc L_{h}$
and $\mc H_{w}$ operate separately, one obtains that the heat and
work currents are\footnote{because of the property $\bra{H_{0}}\mc H_{0}=0$, $\bra{H_{0}}$
is not modified by the transformation to the interaction picture. } $j_{c(h)}=\braOket{H_{0}}{\mc L_{c(h)}}{\tilde{\rho}_{s}(t)}$ and
$j_{w}=\braOket{H_{0}}{\frac{1}{2}\mc H_{w}}{\tilde{\rho}_{s}(t)}$.
In the continuous engine the steady state satisfies $\ket{\tilde{\rho}_{s}(t)}=\ket{\tilde{\rho}_{s}}$
so the total heat and work in steady state in one cycle are:
\begin{eqnarray}
W{}^{\text{cont}} & = & \braOket{H_{0}}{\frac{1}{2}\mc H_{w}}{\tilde{\rho}_{s}}\tau_{cyc},\label{eq: Wcont}\\
Q_{c(h)}^{\text{cont}} & = & \braOket{H_{0}}{\mc L_{c(h)}}{\tilde{\rho}_{s}}\tau_{cyc}.
\end{eqnarray}
These quantities should be compared to the work and heat in the four-stroke
engine. Instead of carrying out the explicit calculation for this
specific four-stroke splitting we use the symmetric rearrangement
theorem (SRT) derived in Appendix III. Symmetric rearrangement of
a Hamiltonian means is a change in the order of couplings $\epsilon(t),\gamma_{c}(t),\gamma_{h}(t)$
that satisfies $\intop\epsilon(t)dt=\text{const},\:\intop\gamma_{c}(t)dt=\text{const},\:\intop\gamma_{h}(t)dt=\text{const}$
and $\epsilon(t)=\epsilon(-t),\gamma_{c}(t)=\gamma_{c}(-t),\gamma_{c}(t)=\gamma_{c}(-t)$.
$\mc H^{II\:stroke}(t)$, $\mc H^{IV\:stroke}(t)$ and any other super
Hamiltonian obtained using the Strang splitting of the continuous
engine, are examples of symmetric rearrangements. The SRT exploits
the symmetry of the Hamiltonian to show that symmetric rearrangement
change heat and work only in $O(s^{3}$). In the Appendix III we show
that 
\begin{eqnarray}
W^{\text{4 stroke}} & \cong & W{}^{\text{cont}},\label{eq: W4  Wc}\\
Q_{c(h)}^{\text{4 stroke}} & \cong & Q_{c(h)}^{\text{cont}}.\label{eq: Q4  Qc}
\end{eqnarray}
Thus, we conclude that up to $s^{3}$ corrections, the engines are
thermodynamically equivalent. When $s\ll1$. work, power heat and
efficiency converge to the same value for all engine types . Clearly,
inside the cycle the work and heat in the two engine are significantly
different ($O(s^{1})$) but after a complete cycle they become equivalent.
The symmetry makes this equivalent more accurate as it holds up to
$s^{3}$ (rather $s^{2}$). Interestingly the work done in the first
half of the cycle is $\frac{1}{2}W{}^{\text{cont}}+O(s^{2})$. However,
when the contribution of the other half is added added the $s^{2}$
correction cancels out and (\ref{eq: W4  Wc}) is obtained (see Appendix
III).

\begin{figure}
\begin{centering}
\includegraphics[width=8.6cm]{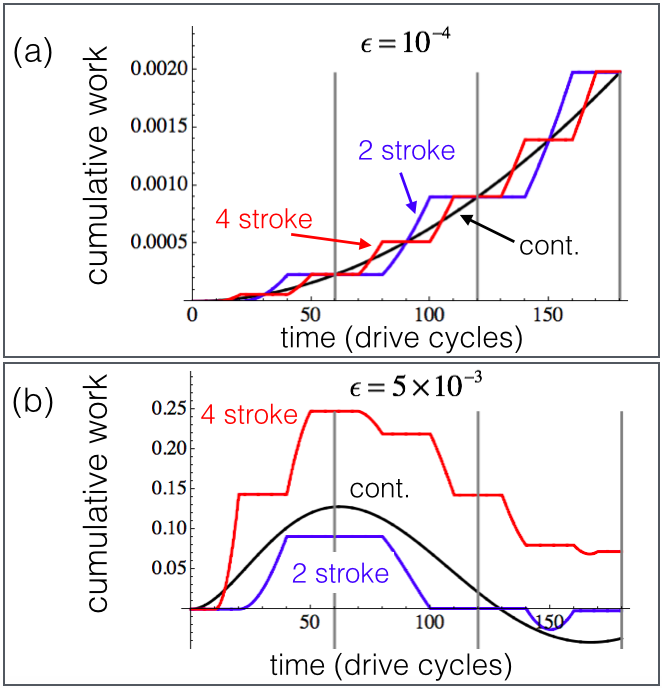}
\par\end{centering}

\label{fig6}\protect\caption{The equivalence of heat engine types in transient evolution when the
engine action is small compared to $\hbar$. (a) The cumulative power
transferred to the work repository is plotted as a function of time.
All engines start in the excited state $\protect\ket 4$, which is
very far from the steady state of the system. At complete engine cycles
(vertical lines) the power in all engines is the same. (b) Once the
action is increased (here the field $\epsilon$ was increased), the
equivalence no longer holds. }

\end{figure}

We emphasize that the SRT and its implications (\ref{eq: W4  Wc}),(\ref{eq: Q4  Qc})
are valid for transients and for any initial state - not just for
steady state operation. In Fig. 6a we show the cumulative work as
a function of time for a four-stroke engine and a continuous engine.
The vertical lines indicate complete cycle of the four-stroke engine.
In addition to the parameter common to all examples specified before,
we used $\epsilon=\gamma_{c}=\gamma_{h}=10^{-4}$ and the equivalence
of work at the vertical lines is apparent. In Fig. 6b the field and
thermal coupling were increased to $\epsilon=\gamma_{c}=\gamma_{h}=5\times10^{-3}$
. Now the engines perform differently even at the end of each cycle.
This example is a somewhat extreme situation where the system changes
quite rapidly (consequence of the initial state we chose). In other
cases, such as steady state operation, the equivalence can be observed
for much larger action values.

The splitting used in (\ref{eq: K 4s}) was based on first splitting
$\mc L_{c}$ and then $\mc H_{w}$. Other engines can be obtained
by different splitting of $\tilde{\mc K}^{\text{cont}}$. For example,
consider the two-stroke engine obtained by splitting $\mc L_{c}+\mc L_{h}$:

\begin{eqnarray}
\tilde{\mc K}^{\text{2 stroke}} & = & e^{-i\frac{3}{2}(\mc L_{c}+\mc L_{h})\frac{\tau_{cyc}}{3}}e^{-i(\frac{3}{2}\mc H_{w})\frac{\tau_{cyc}}{3}}e^{-i\frac{3}{2}(\mc L_{c}+\mc L_{h})\frac{\tau_{cyc}}{3}}.\nonumber \\
\label{eq: K 2stroke}
\end{eqnarray}
Note that in the two-stroke engine the thermal coupling has to be
$\frac{3}{2}$ stronger compared to the continuous case in order to
provide the same action. Using the SRT we obtain the complete equivalence
relations of the three main engine types\footnote{Note that since $\mc K=e^{-i\mc H_{0}\tau_{cyc}}\tilde{\mc K}$, the
equivalence of the evolution operators holds also in the original
frame not just in the interaction frame.}:
\begin{eqnarray}
W^{\text{2 stroke}} & \cong W^{\text{4 stroke}} & \cong W{}^{\text{cont}},\\
Q_{c(h)}^{\text{2 stroke}}\cong & Q_{c(h)}^{\text{4 stroke}}\cong & Q_{c(h)}^{\text{cont}},\\
\tilde{\mc K}^{\text{2 stroke}} & \cong\tilde{\mc K}^{\text{4 stroke}} & \cong\tilde{\mc K}^{\text{cont}}.
\end{eqnarray}

Another type of engine exits when the interaction with the work repository
is carried out by two physically distinct couplings. This happens
naturally if $E_{4}-E_{3}\neq E_{2}-E_{1}$ so that two different
driving lasers have to be used and the Hamiltonian is $H_{0}+\cos((E_{2}-E_{1})t)H_{w1}+\cos((E_{4}-E_{3})t)H_{w2}$.
In such cases, one can make the splitting shown in Fig. 5d. In this
numerical example we use: $H_{w1}=\epsilon(t)\ketbra 12+h.c.$ and
$H_{w2}=\epsilon(t)\ketbra 34+h.c.$ Since there are two different
work strokes in addition to the thermal stroke this engine constitute
a four-stroke engine which differece.

\subsection{Power and energy flow balance}

The average power and heat flow in the equivalence regime are independent
of the cycle time:
\begin{eqnarray}
P_{W} & = & \frac{W}{\tau_{cyc}}=\braOket{H_{0}}{\frac{1}{2}\mc H_{w}}{\tilde{\rho}_{s}},\\
J_{c(h)} & = & \frac{Q_{c(h)}}{\tau_{cyc}}=\braOket{H_{0}}{\mc L_{c(h)}}{\tilde{\rho}_{s}}.
\end{eqnarray}
Using the steady state definition (\ref{eq: steady state}) one obtains
the steady state energy balance equation:
\begin{equation}
P_{w}+J_{c}+J_{h}=0.\label{eq: steady state energy}
\end{equation}
Equation (\ref{eq: steady state energy}) does not necessarily hold
if the system is not in steady state as energy may be temporarily
stored in the baths or in the work repository.

\begin{figure}
\begin{centering}
\includegraphics[width=8.6cm]{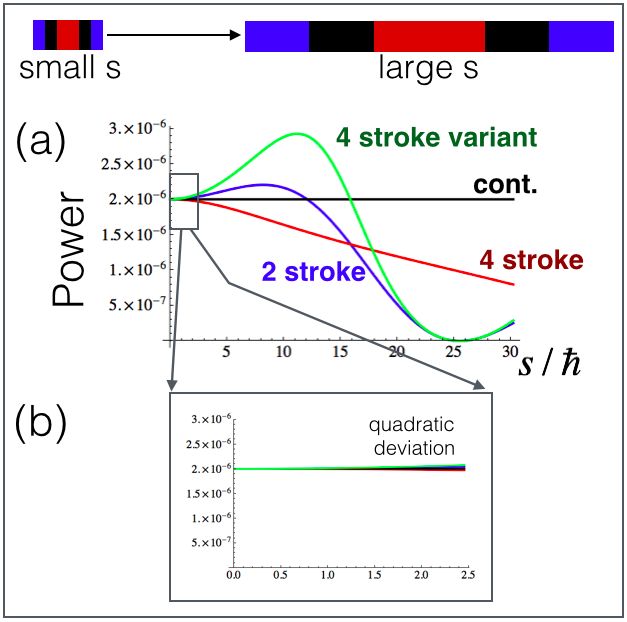}
\par\end{centering}

\label{fig7}\protect\caption{Power as a function of action for various engine types in steady state.
The 4 stroke variant (green) is described at the end of Sec. \ref{sub: Thermo equiv}.
The action is increased by increasing the strokes duration (top illustration).
(a) For large action with respect to $\hbar$, the engines significantly
differ in performance. In this example all engines have the same efficiency,
but they extract different amounts of heat from the hot bath. (b)
In the equivalence regime where the action is small, all engine types
exhibit the same power and also the same heat flows. The condition
$s<\hbar/2$ that follows from the Strang decomposition agrees with
the observed regime of equivalence. The time symmetric structure of
the engines causes the deviation from equivalence to be quadratic
in the action.}
\end{figure}

Figure 7 shows the power in steady state as a function of the action.
The action is increased by increasing the time duration of each stroke
(see top illustration in Fig. 7) . The field and the thermal coupling
are $\epsilon=\gamma_{h}=\gamma_{c}=5\times10^{-4}.$ The coupling
strengths to the bath and work repository are not changed. When the
engine action is large compared to $\hbar$, the engines behave very
differently (Fig. 7a). On the other hand, in the equivalence regime,
where $s$ is small with respect to $\hbar$, the power of all engines
types converges to the same value. In the equivalence regime the power
rises quadratically with the action since the correction to the power
is $s^{3}/\tau_{cyc}\propto\tau_{cyc}^{2}$. This power plateau in
the equivalence regime is a manifestation of quantum interference
effects (coherence in the density matrix), as will be further discussed
in the next section.

The behavior of different engines for large action with respect to
$\hbar$ is very rich and strongly depends on the ratio between the
field and the baths coupling strength. For example, if the field is
amplified then for some parameter the four-stroke engine can produce
more power than the continuous and two stroke engine. Some features
of this diverse dynamics will be discussed elsewhere. 

Finally we comment that the same formalism and results can be extended
for the case the drive that is slightly detuned from the gap.

\subsection{Lasing condition via the equivalence to two-stroke engine}

Laser medium can thought of as continuous engine where the power output
is light amplification. It is well known that lasing requires population
inversion. Scovil et. al \cite{scovil59} were the first to show the
relation between population inversion lasing condition, and the Carnot
efficiency. 

Using the equivalence principle, presented here, the most general
form of the lasing condition can be obtained without any reference
to light-matter interaction. 

Let us start by decomposing the continuous engine into an equivalent
two-stroke engine. For simplicity, it is assumed that the hot and
cold manifolds have some overlap so that in the absence of the driving
field this bath leads the system to a unique steady state $\rho_{0}$.
If the driving field is tiny with respect to the thermalization rates
then the system will be very close to $\rho_{0}$ in steady state.

To see when $\rho_{0}$ can be used for work extraction we need to
discuss passive states. A passive state is a state that is diagonal
in the energy basis, and with populations that decreases monotonically
with the energy \cite{alahverdyan04}. The energy of a passive state
cannot be decreased (or work cannot be extracted from the system)
by applying some unitary transformation (the Hamiltonian after the
transformation is the same as it was before the transformation) \cite{alahverdyan04,alicki13}.
Thus, if $\rho_{0}$ is passive, work cannot be extracted from the
device regardless of the details of the driving field (as long as
it is weak and the equivalence holds). \\
A combination of thermal baths will lead to an energy diagonal $\rho_{0}$.
Consequently, to enable work extraction, passivity must be broken
by population inversion. Therefore, we obtain the standard population
inversion condition. Note that the derivation does not require Einstein
rate equation and any information on the processes of emission and
absorption of photons. 

Furthermore, it now becomes clear that if ``coherent baths'' are
used \cite{scully03} so that $\rho_{0}$ is no longer diagonal in
the energy basis (and therefore no longer passive) it is possible
to extract work even without population inversion. 

In conclusion, using the equivalence principle it is possible to import
known results from work extraction in stroke schemes to continuous
machines.

\section{Quantum thermodynamic signature}

Can thermodynamic measurement reveal quantum effects in the engine?
To answer this we first need to define the corresponding classical
engine.

The term ``classical'' engine is rather ambiguous. There are different
protocols of modifying the system so that is behaves classically.
To make a fair comparison to the fully quantum engine we look for
the minimal modification that satisfies the following conditions:
\begin{enumerate}
\item The dynamics of the device should be fully described using population
dynamics (no coherences, no entanglement).
\item The modification should not alter the energy levels of the system,
the couplings to the baths, and the coupling to the work repository.
\item The modification should not introduce a new source of heat or work.
\end{enumerate}
To satisfy the first requirement we introduce a dephasing operator
that eliminate the coherences\footnote{For simplicity we think of a single particle engine. Thus entanglement
and spin statistics are irrelevant quantum effects. In addition, in
the weak system-bath coupling limit the entanglement to the baths
is negligible.} and leads to a stochastic description description of the engine.
Clearly, a dephasing operators satisfy the second requirement. To
satisfy the third requirement we require ``pure dephasing''; a dephasing
in the energy basis. The populations in the energy basis are invariant
to this dephasing operation. Such a natural source of energy basis
dephasing emerges if there is some scheduling noise \cite{k215}.
That is, if there is some error in the switching time of the strokes.

Let us define a ``quantum-thermodynamic signature'' as a signal
that is impossible to produce by the corresponding classical engine
as defined above. 

Our goal is to derive a threshold for power output that a stochastic
engine cannot exceed but a coherent quantum engine can.

\begin{figure}
\begin{centering}
\includegraphics[width=8.6cm]{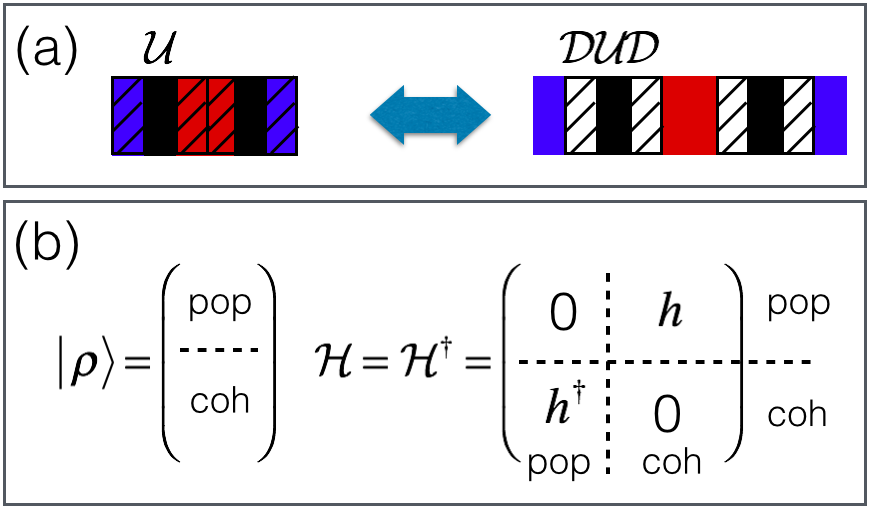}
\par\end{centering}

\protect\caption{\label{fig8}(a left) Dephasing operations (slanted line, operator
$\protect\mc D$) commute with thermal baths so the dephased engine
in (a) left, is equivalent to the one on the right. In the new engine
the unitary evolution is replace by $\protect\mc{DUD}$. If $\protect\mc D$
eliminates all coherences, the effect of $\protect\mc{DUD}$ on the
populations can always be written as a doubly stochastic operator.
(b) Any Hermitian Hamiltonian in Liouville space has the structure
shown in (b). Thus, first order changes in populations critically
depend on the existence of coherence. }

\end{figure}

Before analyzing the effect of decoherence it is instructive to distinguish
between two different work extraction mechanisms in stroke engines.

\subsection{Coherent and stochastic work extraction mechanisms}

Let us consider the work done in the work stroke of a two-stroke engine
(as in Fig. 5c):

\[
W=\braOket{H_{0}}{e^{-i\frac{1}{2}\mc H_{w}\tau_{w}}}{\tilde{\rho}}.
\]
Writing the state as a sum of population and coherences $\ket{\tilde{\rho}}=\ket{\tilde{\rho}_{pop}}+\ket{\tilde{\rho}_{coh}}$
we get:
\begin{eqnarray}
W & = & \braOket{H_{0}}{\sum_{n=1}\frac{(-i\frac{1}{2}\mc H_{w}\tau_{w})^{2n-1}}{(2n-1)!}}{\tilde{\rho}_{coh}}\nonumber \\
 & + & \braOket{H_{0}}{\sum_{n=1}\frac{(-i\frac{1}{2}\mc H_{w}\tau_{w})^{2n}}{(2n)!}}{\tilde{\rho}_{pop}}.\label{eq: P 2 mechanisms}
\end{eqnarray}
This result follows from the generic structure of Hamiltonians in
Liouville space. Any $\mc H$ that originates from an Hermitian Hamiltonian
in Hilbert space (in contrast to Lindblad operators as a source) has
the structure shown in Fig 8b (see Appendix II for Liouville space
derivation of this property). That is, it connects only populations
to coherences and vice versa, but it cannot connect populations to
populations directly\footnote{This is very well known in the context of the Zeno effect.}.
In addition, since $\bra{H_{0}}$ acts as a projection on population
space one gets that odd powers of $\mc H_{w}$ can only operate on
coherences and even powers can only operate on populations. Thus,
the power can be extracted using two different mechanisms. A coherent
mechanism that operates on coherences and a stochastic mechanism that
operates on populations. 

The effect of the ``stochastic'' terms $\sum_{n=1}\frac{(-i\frac{1}{2}\mc H_{w}\tau_{w})^{2n}}{(2n)!}$
on the populations are equivalently described by a single doubly stochastic
operator. If there are no coherences (next section) this leads to
a simple interpretation in terms of full swap events that take place
with some probability.

Continuous engines, on the other hand, have only coherent work extraction
mechanism. This can be seen from the expression for their work output
\begin{equation}
P{}^{\text{cont}}=\braOket{H_{0}}{\frac{1}{2}\mc H_{w}}{\tilde{\rho}}=\braOket{H_{0}}{\frac{1}{2}\mc H_{w}}{\tilde{\rho}_{coh}},\label{eq: Pcont}
\end{equation}
where again we used the population projection property of $\bra{H_{0}}$,
and the structure of $\mc H_{w}$ (Fig. 8b). We conclude that in contrast
to stroke engines, continuous engines have no stochastic work extraction
mechanism. This difference stems from the fact that the in steady
state the state is stationary in continuous engines. Consequently,
there are no higher order terms that can give rise to a population-population
stochastic work extraction mechanism. This is a fundamental difference
between stroke engines and continuous engines. This effect is pronounce
outside the equivalence regime where the stochastic terms become important
(see Sec. \ref{sub: Optimal-thermal-coupling}).

\subsection{Engines subjected to pure dephasing}

Consider the engine shown in Fig 8a. The slanted lines on the baths
indicate that there is an additional dephasing mechanism that takes
place in parallel to the thermalization\footnote{In the Lindblad framework any thermalization is intrinsically associated
with some dephasing. Yet, here we assume an additional controllable
dephasing mechanism.}. Let us denote the evolution operator of the pure dephasing by $\mc D$.
In principle, to analyze the deviation from the coherent quantum engine,
first the steady state has to be solved and then work and heat can
be compared. Even for simple systems this is difficult task. Hence,
we shall take a different approach and derive a power upper bound
for stochastic engines. It is important that the bound will contain
only quantities that are unaffected by the level of coherence in the
system. For example, average energy or dipole expectation values,
do contain information on the coherence. We construct a bound in terms
of the parameters of the system (e.g. the energy levels, coupling
strengths, etc.), an is independent of the state of the system. In
the pure dephasing stage the energy does not change. Hence, the total
energy change in the $\mc{DUD}$ stage is associated with work. 

Let $\mc D_{comp}=\ketbra{pop}{pop}$ be a projection operator on
the population space. This operator generates a complete dephasing
that eliminates all coherences. In such case, the leading order in
the work expression becomes
\begin{eqnarray}
W & = & \braOket{H_{0}}{\mc D_{comp}e^{-i\frac{1}{2}\mc H_{w}\tau_{w}}\mc D_{comp}}{\tilde{\rho}}\nonumber \\
 & = & \frac{\tau_{w}^{2}}{8}\braOket{H_{0}}{\mc H_{w}^{2}}{\tilde{\rho}_{pop}}+O(s^{4}),
\end{eqnarray}
where we used $\bra{H_{0}}\mc D=\bra{H_{0}}$ and $\mc D_{comp}\ket{\tilde{\rho}}=\ket{\tilde{\rho}_{pop}}$.
Since $\mc D_{comp}$ eliminates coherences, $W$ does not contain
a linear term in time. Next by using the following relation: $\text{\ensuremath{\braOket{H_{0}}B{\rho}}}\le\sqrt{\braket{H_{0}}{H_{0}}\braket{\rho}{\rho}}\nrm B$,
$\sqrt{\braket{H_{0}}{H_{0}}}=\sqrt{tr(H_{0}^{2})}$, we find that
for $s\ll\hbar$ the power of a stochastic engine satisfies:
\begin{eqnarray}
P_{stoch} & \le & \frac{z}{8}\sqrt{tr(H_{0}^{2})-tr(H_{0})^{2}}\Delta_{w}^{2}d^{2}\tau_{cyc},\label{eq: power signature}\\
z=1 &  & \text{two-stroke}\nonumber \\
z=1/2 &  & \text{four-stroke}\nonumber 
\end{eqnarray}
where $\Delta_{w}$ is the gap of the interaction Hamiltonian (maximal
eigenvalue minus minimal eigenvalue of $H_{w}$), and $d$ is the
duty cycle - the fraction of time dedicated to work extraction (e.g.
$d=1/3$ in all the examples in this paper). We also used the fact
that $\braket{\rho_{pop}}{\rho_{pop}}$ is always smaller than the
purity $\braket{\rho}{\rho}$ and therefore smaller than one. Note
that, as we required, this bound is state-independent, and the right
hand side of (\ref{eq: power signature}) contains no information
on the coherences in the system. As shown earlier, in coherent quantum
engines (in the equivalence regime) the work scales linearly with
$\tau_{cyc}$ (see (\ref{eq: Wcont}) and (\ref{eq: W4  Wc})) and
therefore the power is constant as a function of $\tau_{cyc}$. When
there are no coherences the power scales linearly with $\tau_{cyc}$. 

Numerical results of power as function of cycle time are shown in
Fig. 9. The power is not plotted as function of action as before,
because at the same cycle time the coherent engine and dephased engine
have different action. The action of the dephased engine is
\begin{equation}
s_{deph}=(\nrm{\mc L_{c}}+\nrm{\mc L_{h}}+\nrm{\frac{1}{2}\mc H_{w}}+\nrm{\mc L_{dephasing}})\tau_{cyc}.
\end{equation}
If the dephasing is significant the action is large and equivalence
cannot be observed. That is, a fully stochastic engine in quantum
system have large action and cannot satisfy $s\ll\hbar$.

The chosen coherence time is $100\tau_{d}$. When the cycle time is
small with respect to the coherence time, equivalence is observed.
Yet, the power is significantly smaller compared to the fully coherent
case. For longer cycles the decoherence starts to take effect, and
the expected linear power growth is observed. The stochastic power
bounds for a two-stroke engine (dashed-blue), and for a four-stroke
engine (dashed-red) define a power regime (shaded area) that is inaccessible
to fully stochastic engines. Thus, any power measurement in this regime
unequivocally indicates the presence of quantum coherences in the
engine. Note, that to measure power the work repository is measured
and not the engine. Furthermore, the engine must operate for many
cycles to reduce fluctuations in the accumulated work. To calculate
the average power the accumulated work is divided by the total operation
time and compared to the stochastic power threshold (\ref{eq: power signature}). 

Note that had we chosen complete dephasing then the power output of
the continuous engine would have been zero as expected from (\ref{eq: Pcont}).

In summary, quantum thermodynamics signature in stroke engines can
be observed in the weak action limit.

\section{\label{sub: Optimal-thermal-coupling}The over thermalization effect
in coherent quantum heat engine}

In all the numerical examples studied so far, the unitary action and
the thermal action were roughly comparable for reason that will soon
become clear. In this section we study some generic features that
take place when the thermal action takes over.

\begin{figure}
\begin{centering}
\includegraphics[width=8.6cm]{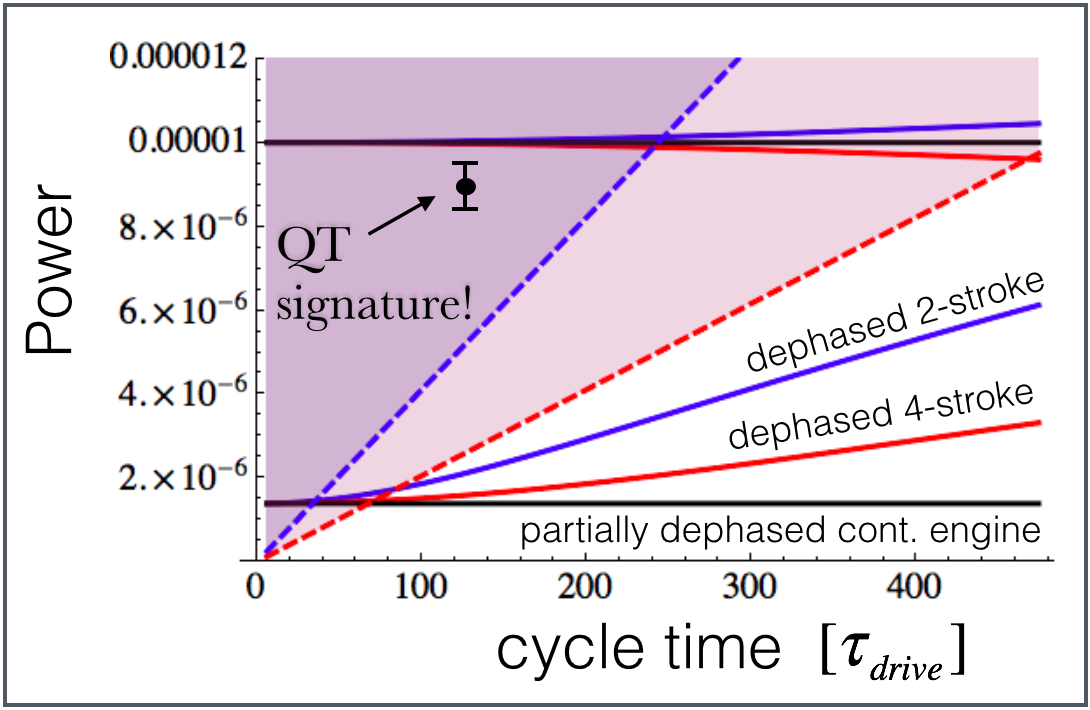}
\par\end{centering}

\label{fig9}\protect\caption{The output of the three types of engines (two-stroke blue, four stroke
red, continuous black) with and without dephasing (same as Fig. 7b).
The dephasing time is $100\tau_{drive}$. Well above the dephasing
time ($\sim200\tau_{drive}$) the power grows linearly as expected
from stochastic engines (three bottom solid lines). Below the dephasing
time ($\sim20\tau_{drive}$) equivalence is observed. Yet the power
is significantly lower compared to the coherent engine (top three
solid lines). The dashed lines show the stochastic upper bound on
the power for two-stroke (dashed-blue) and four-stroke (dashed-red)
engines. Any power measurement in the shaded area of each engine indicates
the presence of quantum interference in the engine. This plot also
demonstrates that for weak couplings (low action) coherent engines
produce much more power compared to stochastic dephased engines. }
\end{figure}

Let us now consider the case where the unitary contribution to the
action $\nrm{\mc H_{\omega}}\tau$ is small with respect to $\hbar$.
All the time intervals are fixed but we can control the thermalization
rate $\gamma$ (for simplicity, we assume it is the same value for
both baths). Common sense suggests that increasing $\gamma$ should
increase the power output. At some stage this increase will stop since
the system will already reach thermal equilibrium with the bath (or
baths in two-stroke engines). Yet, Fig. 10 shows that there is a very
distinctive peak where an optimal coupling takes place. That is, in
some cases less thermalization leads to more power. We call this effect
over thermalization. This effect is generic and not unique to the
specific model used in the numerical simulations. The parameter used
for the plot are $\epsilon=\gamma_{c}=\gamma_{h}=2\times10^{-4}$
and the number of drive cycle pen engine cycle is $m=600$.

\begin{figure}
\begin{centering}
\includegraphics[width=8.6cm]{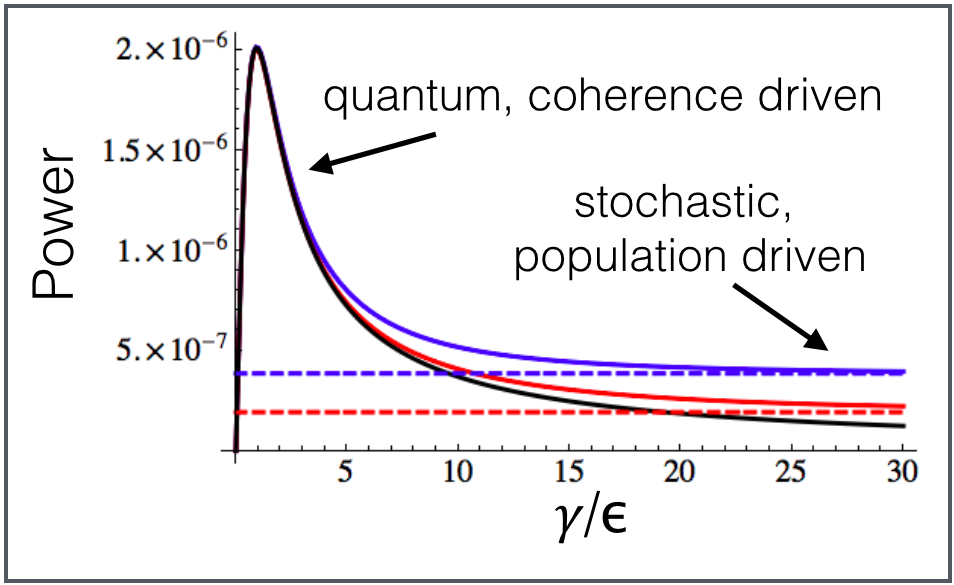}
\par\end{centering}

\label{fig10}\protect\caption{The over thermalization effect is the decrease of power when the thermalization
rate is increased. Over thermalization degrades the coherent work
extraction mechanism without effecting the stochastic work extraction
mechanism. When the coherent mechanism gets weak enough, the power
is dominated by the stochastic power extraction mechanisms and power
saturation is observed (dashed lines). The continuous engine has no
stochastic work extraction mechanism and therefore it decays to zero
without reaching saturation. }

\end{figure}

The peak is a consequence of the interplay between the two different
work extraction mechanisms (see Sec. 4.1). For low $\gamma$ the coherences
in the system are significant and the leading term in the power is
$\braOket{H_{0}}{-i\frac{1}{2}\mc H_{w}}{\tilde{\rho}_{coh}}d$ (where
d is the duty cycle). In principle, all Lindblad thermalization processes
are associated with some level of decoherence. This decoherence generates
an exponential decay of $\ket{\tilde{\rho}_{coh}}$ that explains
the decay on the right hand side of the peak. At a certain stage the
linear term becomes so small that the stochastic second order term
$-\frac{1}{8}\braOket{H_{0}}{\mc H_{w}^{2}}{\tilde{\rho}_{pop}}d^{2}\tau_{cyc}$
dominates the power. $\ket{\tilde{\rho}_{pop}}$ eventually saturates
for large $\gamma$ and therefore the stochastic second order term
leads to a power saturation. Interestingly, in the example shown in
Fig. (\ref{fig10}) we observe that the peak is obtained when $\gamma$
and $\epsilon$ are roughly equal. Of course, what really matters
is the thermal action with respect to unitary action and not just
the values of the parameter $\gamma$ and $\epsilon$. 

If thermalization occurs faster, the thermal stroke can be shortened
and this increases the power. However, this effect is small with respect
to the exponential decay of the coherences. We conclude that even
without additional dephasing as in the previous section, excessive
thermal coupling turns the engine into a stochastic machine. For small
unitary action this effect severely degrades the power output. The
arguments presented here are valid for any small action coherent quantum
engine.

\section{Concluding remarks}

We identified coherent and stochastic work extraction mechanisms in
quantum heat engines. While stroke engines have both, continuous engines
only have the coherent mechanism. We introduced the ``norm action''
of the engine using Liouville space and showed that when this action
is small compared to $\hbar$ all three engine types are equivalent.
This equivalence emerges because for small action only the coherent
mechanism is important. Despite the equivalence, before the engine
cycle is completed the state of the different engine type differ by
$O(s^{1})$. This holds true also for work and heat. Remarkably, at
the end of each engine cycle a much more accurate $O(s^{3})$ equivalence
emerges. Furthermore, the equivalence holds also for transient dynamics,
even when the initial state is very far from the steady state of the
engine. It was shown that for small action the coherent work extraction
is considerably stronger than the stochastic work extraction mechanism.
This enabled us to derive a power bound for stochastic engines that
constitutes a quantum thermodynamics signature. Any power measurement
that exceeds this bound indicated the presence of quantum coherence
and the operation of the coherent work extraction mechanism.

The present derivation makes no assumption on the direction of heat
flows and the sign of work. Thus our result are equally applicable
to refrigerators and heaters. 

It is interesting to try and apply these concepts of equivalence and
quantum thermodynamic signatures to more general scenarios: non Markovian
baths, engines with non symmetric unit cell, and engines with correlation
between different particles (entanglement and quantum discord). We
conjecture that in multiple particle engines entanglement will play
a similar role to that of coherence in single particle engines.

Work support by the Israeli science foundation. Part of this work
was supported by the COST Action MP1209 'Thermodynamics in the quantum
regime'.

\section*{Appendix I - Strang decomposition validity}

Let $\mc K$ be an operator generated by two non commuting operators
$\mc A$ and $\mc B$:
\[
\mc K=e^{(\mc A+\mc B)dt}.
\]
The splitted operator is

\[
\mc K_{s}=e^{\frac{1}{2}\mc Adt}e^{\mc Bdt}e^{\frac{1}{2}\mc Adt}.
\]
Our goal is to quantify the difference between $\mc K$ and $\mc K_{s}$,
$\nrm{\mc K_{s}-\mc K}$ where $\nrm{\cdot}$ stands for the spectral
norm. In principle, other sub-multiplicative matrix norm can be used
(such as the Hilbert Schmidt norm). However, the spectral norm captures
more accurately aspects of quantum dynamics \cite{ruPuritySpeed,uzdin100evoSpeed,uzdinEmbbeding,uzdinNHresources}.
$\mc K$ can be expanded as:
\begin{equation}
\mc K=\sum\frac{(\mc A+\mc B)^{n}dt^{n}}{n!}.
\end{equation}
$\mc K_{s}$ on the other hand is:
\begin{eqnarray}
\mc K_{s} & = & \sum_{k,l,m=0}^{\infty}\frac{(\mc A/2)^{k}dt^{k}}{k!}\frac{\mc B^{l}dt^{l}}{l!}\frac{(\mc A/2)^{m}dt^{m}}{m!}\nonumber \\
 & = & \sum_{n=0}^{\infty}\sum_{l=0}^{n}\sum_{k=0}^{n-l}\frac{(\mc A/2)^{k}}{k!}\frac{\mc B^{l}}{l!}\frac{(\mc A/2)^{n-l-k}}{(n-l-k)!}dt^{n}.
\end{eqnarray}
Due to the symmetric splitting the terms up to $n=2$ (including n=2)
are identical for both operators. Therefore the difference can be
written as 
\begin{eqnarray}
\nrm{\mc K_{s}-\mc K} & =\nonumber \\
 &  & \left\Vert \sum_{n=3}^{\infty}\sum_{l=0}^{n}\sum_{k=0}^{n-l}\frac{(\mc A/2)^{k}}{k!}\frac{\mc B^{l}}{l!}\frac{(\mc A/2)^{n-l-k}}{(n-l-k)!}dt^{n}\right.\nonumber \\
 & - & \left.\sum_{n=3}\frac{(\mc A+\mc B)^{n}dt^{n}}{n!}\right\Vert .
\end{eqnarray}
next we apply the triangle inequality and the sub-multiplicativity
property and get:
\begin{eqnarray}
\nrm{\mc K_{s}-\mc K} & \le\nonumber \\
 &  & \left\Vert \sum_{n=3}^{\infty}\sum_{l=0}^{n}\sum_{k=0}^{n-l}\frac{\nrm{\mc A/2}{}^{k}}{k!}\frac{\nrm B^{l}}{l!}\frac{\nrm{\mc A/2}{}^{n-l-k}}{(n-l-k)!}dt^{n}\right.\nonumber \\
 &  & \left.+\sum_{n=3}^{\infty}\frac{(\nrm{\mc A}+\nrm{\mc B})^{n}dt^{n}}{n!}\right\Vert .
\end{eqnarray}
Using the binomial formula two times one finds
\begin{eqnarray}
\sum_{n=3}^{\infty}\sum_{l=0}^{n}\sum_{k=0}^{n-l}\frac{\nrm{\mc A/2}{}^{k}}{k!}\frac{\nrm{\mc B}^{l}}{l!}\frac{\nrm{\mc A/2}{}^{n-l-k}}{(n-l-k)!}dt^{n} & =\nonumber \\
\sum_{n=3}^{\infty}\frac{(\nrm{\mc A}+\nrm{\mc B})^{n}dt^{n}}{n!},
\end{eqnarray}
and therefore
\begin{eqnarray}
\nrm{\mc K_{s}-\mc K} & \le & 2\sum_{n=3}^{\infty}\frac{(\nrm{\mc A}+\nrm{\mc B})^{n}dt^{n}}{n!}\nonumber \\
 & = & 2R_{2}[(\nrm{\mc A}+\nrm{\mc B})dt].
\end{eqnarray}
The right hand side is the Taylor reminder of a power series of an
exponential with as argument $s=(\nrm{\mc A}+\nrm{\mc B})dt$. The
Taylor reminder formula for the exponent function is $R_{k}(x)=e^{\xi}\frac{\left|s\right|^{k+1}}{(k+1)!}$
where $0\le\xi\le1$ (for now we assume $s<1$). Setting $k=2$ and
$\xi=1$ (worst case), we finally obtain
\begin{eqnarray}
\nrm{\mc K_{s}-\mc K} & \le & \frac{e}{3}[(\nrm A+\nrm{\mc B})dt]^{3}\le s^{3},\\
s & = & (\nrm A+\nrm{\mc B})dt.
\end{eqnarray}

To get an estimation where the leading non neglected term of $\mc K$,
$(\mc A+\mc B)^{2}dt^{2}/2$ is larger then the reminder we require
that

\begin{eqnarray}
(\nrm{\mc A+\mc B})^{2}dt^{2}/2 & \ge & s^{3}.
\end{eqnarray}
Using the triangle inequality we get the estimated condition for the
Strang decomposition: 
\begin{equation}
s\le1/2.
\end{equation}
This condition explains why it was legitimate to limit the range of
$s$ to 1 in the reminder formula.

\section*{Appendix II - Liouville space formulation of quantum dynamics}

Quantum dynamics is traditionally described in Hilbert space. However,
it is convenient, in particular for open quantum systems, to introduce
an extended space where density operators are vectors and time evolution
is generated by a Schrödinger-like equation. This space is usually
referred to as Liouville space \cite{mukamel1995principles}. We denote
the \textquotedbl{}density vector\textquotedbl{} by $\ket{\rho}\in\mathbb{C}^{1\times N^{2}}$.
It is obtained by reshaping the density matrix $\rho$ into a larger
single vector with index $\alpha\in\{1,2,....N^{2}\}.$ The one-to-one
mapping of the two matrix indices into a single vector index $\{i,j\}\to\alpha$
is arbitrary, but has to be used consistently. The vector $\ket{\rho}$
is not normalized to unity in general. Its norm is equal to the purity,
$\mc P=\text{tr}(\rho^{2})=\braket{\rho}{\rho}$ where $\bra{\rho}=\ket{\rho}^{\dagger}$
as usual. The equation of motion of the density vector in Liouville
space follows from $d_{t}\rho_{\alpha}=\sum_{\beta}\rho_{\beta}\partial(d_{t}\rho_{\alpha})/\partial\rho_{\beta}$.
Using this equation, one can verify that the dynamics of the density
vector $\ket r$ is governed by a Schrödinger-like equation in the
new space, 
\begin{equation}
i\partial_{t}\ket{\rho}=\mc H\ket{\rho},\label{eq: schrodinger eq}
\end{equation}
where the super-Hamiltonian $\mc H^{tot}\in\mathbb{C}^{N^{2}\times N^{2}}$
is given by, 
\begin{equation}
\mc H_{\alpha\beta}=i\frac{\partial(d_{t}\rho_{\alpha})}{\partial\rho_{\beta}}.\label{eq: Hr form}
\end{equation}

A particularly useful index mapping is described in \cite{machnes14}
and in \cite{roger1994topics}. For this form $\mc H$ has a very
simple form in terms of original Hilbert space Hamiltonian and Lindblad
operators. 

$\mc H=\mc H^{H}+\mc L$ is non-Hermitian for open quantum systems.
$\mc H^{H}$ originates from the Hilbert space Hamiltonian $H$, and
$\mc L$ from the Lindblad terms%
. $\mc H^{H}$ is always Hermitian. The skew-Hermitian part $(\mc L-\mc L^{\dagger})/2$
is responsible for purity changes. Yet, in Liouville space, the Lindblad
operators $A_{k}$ (\ref{eq: Lind Hil}) may also generate an Hermitian
term $(\mc L+\mc L^{\dagger})/2$. Though Hermitian in Liouville space
this term cannot be associated with a Hamiltonian in Hilbert space. 

For time-independent $\mc H$ the evolution operator in Liouville
space is:
\begin{equation}
\ket{\rho(t)}=\mc K\ket{\rho(t')}=e^{-i\mc H(t-t')}\ket{\rho(t')}.
\end{equation}
If $\mc L=0$, $\mc K$ is unitary%
The fact that the evolution operator can be written as an exponent
of a matrix, without any commutators as in Hilbert space, is a very
significant advantage (see for example \cite{ruPuritySpeed}). It
is important to note that not all vectors in Liouville space can be
populated exclusively. This is due to the fact that only positive
$\rho$ with unit trace are legitimate density matrices. The states
that can be populated exclusively describe steady states, while others
correspond to transient changes. We remind that in this paper we will
use calligraphic letters to describe operators in Liouville space
and ordinary letters for operators in Hilbert space. For states, however,
$\ket A$ will denote a vector in Liouville space formed from $A_{N\times N}$
by ``vec-ing'' $A$ into a column in the same procedure $\rho$
is converted into $\ket{\rho}$.

\subsection*{Useful relations in Liouville space.}

In Liouville space, the standard inner product of two operators in
Hilbert space $\text{tr}A^{\dagger}B$ reads
\[
\text{tr}A^{\dagger}B=\braket AB.
\]
In particular the purity $\mc P=\braket rr$ is just the square of
the distance from the origin in Liouville space.

A useful relation for $\mc H^{H}$ :
\begin{equation}
\mc H^{H}\ket H=\bra H\mc H^{H}=0.\label{eq: vec op herm identity}
\end{equation}
The proof is as follows:

\begin{eqnarray}
\mc{\mc H}_{ij,mn}^{H} & = & H_{im}\delta_{jn}-H_{nj}\delta_{im}.\label{eq: Hherm Lio}
\end{eqnarray}
Therefore, using (\ref{eq: Hherm Lio}) we get:
\begin{eqnarray}
\mc H^{H}\ket H=\sum_{\beta}\mc H_{\alpha\beta}^{H}H_{\beta}=\sum_{mn}\mc H_{ijmn}^{H}H_{mn} & = & [H,H]=0\nonumber \\
\end{eqnarray}
This property is highly useful. We stress that (\ref{eq: vec op herm identity})
is a property of Hermitian operators in Hilbert space where both $H$
and $\mc H$ are well defined. A general Hermitian operator in Liouville
space may not have a corresponding $H$ in Hilbert space.

Another property that immediately follows from (\ref{eq: Hherm Lio})
is 
\begin{equation}
\mc{\mc H}_{ii,kk}^{H}=0.
\end{equation}
This corresponds to a well known property of unitary operation. If
the system starts from a diagonal density matrix, then for short times
the evolution generated by $\mc H^{H}$, $e^{-i\mc H^{H}dt}=I-i\mc H^{H}dt+O(dt^{2})$
does not change the population in the leading order.

\subsection*{Expectation values and their time evolution in Liouville space}

The expectation value of an operator in Hilbert space is $\left\langle A\right\rangle =tr(\rho A)$
. Since $\rho$ is Hermitian the expectation value is equal to the
inner product of $A$ and $\rho$ and therefore:
\[
\left\langle A\right\rangle =tr(\rho A)=\braket{\rho}A.
\]
the dynamics of $\left\langle A\right\rangle $ under Lindblad evolution
operator:
\begin{equation}
\frac{d}{dt}\left\langle A\right\rangle =-i\braOket A{\mc H}{\rho}+\braket{\rho}{\frac{d}{dt}A}.\label{eq: Lio expect change}
\end{equation}
\textit{Note that in Liouville space there is no commutator term}
since $\mc H$ operates on $\ket{\rho}$ just from the left. If the
total Hamiltonian is Hermitian and time independent the conservation
of energy follows immediately from applying (\ref{eq: Lio expect change})
and (\ref{eq: vec op herm identity}) for $A=H$.

\section*{Appendix III - The symmetric rearrangement theorem (SRT)}

The goal of this Appendix is to explain why the equivalence of evolution
operators leads to equivalence of work and equivalence of heat. In
addition it is shown why it is valid also for transients. For the
equivalence of evolution operator we have required that the super
Hamiltonian is symmetric and that the action is small:
\begin{eqnarray}
\mc H(t) & = & \mc H(-t),\\
s & = & \int_{-\tau/2}^{+\tau/2}\nrm{\mc H}dt\ll\hbar.
\end{eqnarray}
Let the initial state at time $t=-\tau/2$ be
\begin{equation}
\ket{\tilde{\rho}_{i}}=\ket{\tilde{\rho}(-\tau/2}.
\end{equation}
this state leads to a final state at $\tau/2$
\begin{equation}
\ket{\tilde{\rho}_{f}}=\ket{\tilde{\rho}(\tau/2}.
\end{equation}
Our goal is to evaluate a symmetric expectation value difference of
the form:
\begin{eqnarray}
dA_{tot} & = & [\left\langle A(t_{2})\right\rangle -\left\langle A(t_{1})\right\rangle ]+[\left\langle A(-t_{1})\right\rangle -\left\langle A(-t_{2})\right\rangle ]\nonumber \\
 & = & [\braket A{\tilde{\rho}(t_{2})}-\braket A{\tilde{\rho}(t_{1})}]\nonumber \\
 &  & +[\braket A{\tilde{\rho}(-t_{1})}-\braket A{\tilde{\rho}(-t_{2})}],\\
t_{2},t_{1} & \ge & 0\nonumber 
\end{eqnarray}
that is, the change in the expectation value of $A$ in the segment
$[t_{1},t_{2}]$ and its symmetric counterpart in negative time (e.g.
the green areas in Fig. 11a). When $A$ is equal to $H_{0}$ this
difference will translate into work or heat. We start with the expansion:
\begin{eqnarray}
[\left\langle A(t_{2})\right\rangle -\left\langle A(t_{1})\right\rangle ]=\braket A{\mc K_{t_{1}\to t_{2}}-I|\tilde{\rho}(t_{1})} & =\nonumber \\
\braket A{-i\mc H(t_{1})\delta t-\frac{1}{2}\mc H(t_{1})^{2}\delta t^{2}|\tilde{\rho}(t_{1})}+O(s^{3}).\nonumber \\
\end{eqnarray}
For the negative side we get:
\begin{eqnarray}
[\left\langle A(-t_{1})\right\rangle -\left\langle A(-t_{2})\right\rangle ]=\braket A{I-\mc K_{-t_{1}\to-t_{2}}|r(-t_{1})} & =\nonumber \\
\braket A{-i\mc H(-t_{1})\delta t+\frac{1}{2}\mc H(-t_{1})^{2}\delta t^{2}|\tilde{\rho}(-t_{1})}+O(s^{3}).\nonumber \\
\end{eqnarray}
Next we use the fact that:
\begin{eqnarray}
\ket{\tilde{\rho}(t_{1})} & = & \ket{\tilde{\rho}(0)}-i\int_{0}^{t_{1}}\mc H(t)dt\ket{\tilde{\rho}(0)}+O(s^{2}),\nonumber \\
\label{eq: r(t1) r(t0)}\\
\ket{\tilde{\rho}(-t_{1})} & = & \ket{\tilde{\rho}(0)}+i\int_{0}^{t_{1}}\mc H(t)dt\ket{\tilde{\rho}(0)}+O(s^{2}).\nonumber \\
\label{eq: r(-t1) r(t0)}
\end{eqnarray}
When adding the two segments the second order cancels out and we get:
\begin{equation}
\delta A_{tot}=-2i\braOket A{\mc H(t_{1})}{\tilde{\rho}(0)}\delta t+O(s^{3}).\label{eq: dAtot r(0)}
\end{equation}
Note that the result using $\ket{\tilde{\rho}(0)}$ which is not given
explicitly. To correctly relate it to $\ket{\tilde{\rho}(-\tau/2)}$
we have to use symmetric rearrangement properties of the evolution
operator.

\subsection*{Symmetric rearrangement}

\begin{figure}
\begin{centering}
\includegraphics[width=8.6cm]{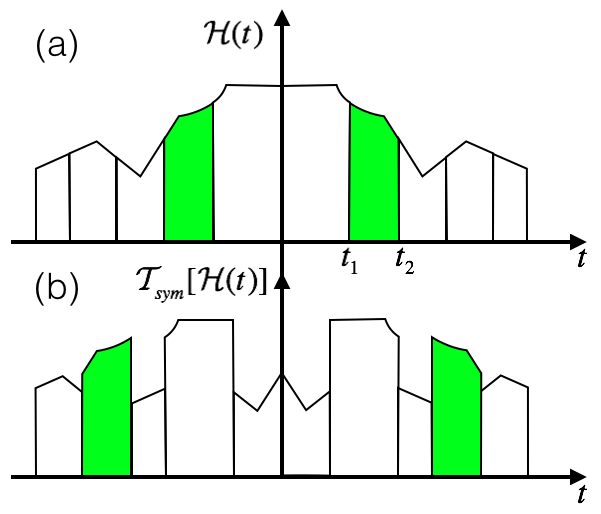}
\par\end{centering}

\protect\caption{The Hamiltonians in (a) and (b) are related by symmetric rearrangement
of the time segments. Up to a small correction $O(s^{3})$ the change
in expectation values of an observable $A$ that takes place during
the green segments is the same in both cases. This effect explains
why work and heat are the same in various types of engines when $s$
is small compared to $\hbar$ (equivalence regime).}
\end{figure}

In Fig. 11a there is an illustration of some time dependent Hamiltonian
with reflection symmetry $\mc H(t)=\mc H(-t)$. We use $\mc H$ to
denote a Liouville space operator which may be any unitary operation
or Markovian Lindblad operation. Assume that in addition to the symmetric
bins of interest (green bins) the reminder of the time is also divided
into bins in a symmetric way so that there is still a reflection symmetry
also in the bin partitioning. Now, we permute the bins in the positive
side as desired, and then make the opposite order in the negative
side so that the reflection symmetry is kept. An example of such an
operation is shown in Fig. 11b. Due to the Strang decomposition we
know that the total evolution operator will stay the same under this
rearrangement up to third order:
\begin{equation}
\mc K_{-\frac{\tau}{2}\to\frac{\tau}{2}}=\mc T_{sym}[\mc K]_{-\frac{\tau}{2}\to\frac{\tau}{2}}+O(s^{3}),\label{eq: Ueq}
\end{equation}
where $\mc T_{sym}[x]$ stands for evaluation of $x$ after a symmetric
reordering.

\subsection*{The symmetric rearrangement theorem (SRT)}

From (\ref{eq: Ueq}) we get that if the initial state is the same
for a system described by $\mc K$, and for a system described by
$\mc T_{sym}[\mc K]$, the final state at $t=\tau/2$ is the same
for both systems up to a third order correction:
\begin{equation}
\ket{\tilde{\rho}(\frac{\tau}{2})}=\mc T_{sym}[\ket{\tilde{\rho}(\frac{\tau}{2})}]+O(s^{3}).\label{eq: rf eq}
\end{equation}
Using (\ref{eq: r(t1) r(t0)}),(\ref{eq: r(-t1) r(t0)}) we see that:
\begin{equation}
\ket{\tilde{\rho}(0)}=\frac{\ket{\tilde{\rho}(\frac{\tau}{2})}+\ket{\tilde{\rho}(-\frac{\tau}{2})}}{2}+O(s^{2}),\label{eq: r(0)}
\end{equation}
and because of (\ref{eq: rf eq}) it also holds that:
\begin{equation}
\mc T_{sym}[\ket{\tilde{\rho}(0)}]=\ket{\tilde{\rho}(0)}+O(s^{2})=\frac{\ket{\tilde{\rho}(\frac{\tau}{2})}+\ket{\tilde{\rho}(-\frac{\tau}{2})}}{2}+O(s^{2}),
\end{equation}
using this in (\ref{eq: dAtot r(0)}) we get that:
\begin{equation}
\delta A_{tot}=-2i\bra A\mc H(t_{1})\frac{\ket{\tilde{\rho}(\frac{\tau}{2})}+\ket{\tilde{\rho}(-\frac{\tau}{2})}}{2}\delta t+O(s^{3}).\label{eq: dA -t/2 t/2}
\end{equation}
Expression (\ref{eq: dA -t/2 t/2}) no longer depends on the position
of the time segment, but only on its duration and on the value of
$\mc H$. Thus, the SRT states that the expression above also holds
for any symmetric rearrangement
\begin{equation}
dA_{tot}=\mc T_{sym}[dA_{tot}]+O(s^{3}).
\end{equation}

If we replace $A$ by $H_{0}$ and $\mc H(t_{1})$ by $\mc L_{c},\mc L_{h}$
or $\mc H_{w}$ we immediately get the invariance of heat and work
to symmetric rearrangement (up to $s^{3}$). If $\ket{\tilde{\rho}(-\frac{\tau}{2})}$
is the same for all engines then $\ket{\tilde{\rho}(\frac{\tau}{2})}$
is also the same for all engines types up to $O(s^{3})$. Consequently
for all stroke engines the expression for work and heat are:
\begin{eqnarray}
W & = & -2i\bra A\intop_{t\in t_{w}}\mc H_{w}(t)dt\frac{\ket{\tilde{\rho}(\frac{\tau}{2})}+\ket{\tilde{\rho}(-\frac{\tau}{2})}}{2}+O(s^{3}),\nonumber \\
\label{eq: W srt}\\
Q_{c(h)} & = & -2i\bra A\intop_{t\in t_{c(h)}}\mc L_{c(h)}(t)dt\frac{\ket{\tilde{\rho}(\frac{\tau}{2})}+\ket{\tilde{\rho}(-\frac{\tau}{2})}}{2}+O(s^{3}).\nonumber \\
\label{eq: Q srt}
\end{eqnarray}

Using the identity $\ket{\tilde{\rho}(\frac{\tau}{2})}+\ket{\tilde{\rho}(-\frac{\tau}{2})}=\ket{\tilde{\rho}(t)}+\ket{\tilde{\rho}(-t)}+O(s^{2})$
that follows from (\ref{eq: r(0)}), the integration over time of
the energy flows $j_{w}=\braOket{H_{0}}{\frac{1}{2}\mc H_{w}}{\tilde{\rho}(t)}$
and $j_{c(h)}=\braOket{H_{0}}{\mc L_{c(h)}}{\tilde{\rho}(t)}$ for
continuous engines, yields expressions (\ref{eq: W srt}) and (\ref{eq: Q srt})
once more. This implies that the SRT (\ref{eq: W srt}) and (\ref{eq: Q srt})
holds even if the different operations $\mc L_{c},\mc L_{h}$ and
$\mc H_{w}$ overlap with each other.

We emphasize that all the above relations hold for any initial state
and not only in steady state where $\ket{\tilde{\rho}(\frac{\tau}{2})}=\ket{\tilde{\rho}(-\frac{\tau}{2})}$.
The physical implication is that in the equivalence regime different
engines are thermodynamically indistinguishable when monitored at
the end of each cycle, even when the system is not in its steady state.

\bibliographystyle{apsrev}
\bibliography{citeamikamronnie1}

\end{document}